\documentclass[aps,pre,reprint,superscriptaddress,amsmath,amssymb]{revtex4-2}

\usepackage{graphicx}
\usepackage{txfonts}
\usepackage[hypertexnames=false]{hyperref}
\usepackage{braket}
\usepackage{bm}

\begin{document}

\title{Universality in the dynamical phase transitions of Brownian motion}
\author{Takahiro Kanazawa}
\affiliation{Department of Physics, The University of Tokyo, 7-3-1 Hongo, Bunkyo-ku, Tokyo 113-0033, Japan}
\affiliation{Nonequilibrium Physics of Living Matter RIKEN Hakubi Research Team, RIKEN Center for Biosystems Dynamics Research, 2-2-3 Minatojima-minamimachi, Chuo-ku, Kobe 650-0047, Japan}
\author{Kyogo Kawaguchi}
\affiliation{Department of Physics, The University of Tokyo, 7-3-1 Hongo, Bunkyo-ku, Tokyo 113-0033, Japan}
\affiliation{Nonequilibrium Physics of Living Matter RIKEN Hakubi Research Team, RIKEN Center for Biosystems Dynamics Research, 2-2-3 Minatojima-minamimachi, Chuo-ku, Kobe 650-0047, Japan}
\affiliation{RIKEN Cluster for Pioneering Research, 2-2-3 Minatojima-minamimachi, Chuo-ku, Kobe 650-0047, Japan}
\affiliation{Institute for Physics of Intelligence, The University of Tokyo, 7-3-1 Hongo, Bunkyo-ku, Tokyo 113-0033, Japan}
\author{Kyosuke Adachi}
\affiliation{Nonequilibrium Physics of Living Matter RIKEN Hakubi Research Team, RIKEN Center for Biosystems Dynamics Research, 2-2-3 Minatojima-minamimachi, Chuo-ku, Kobe 650-0047, Japan}
\affiliation{RIKEN Interdisciplinary Theoretical and Mathematical Sciences Program, 2-1 Hirosawa, Wako 351-0198, Japan}

\date{\today}

\begin{abstract}
We study the dynamical phase transitions (DPTs) appearing for a single Brownian particle without drift.
We first explore how first-order DPTs in large deviations can be found even for a single Brownian particle without any force upon raising the dimension to higher than four.
The DPTs accompany temporal phase separations in their dynamical paths, which we numerically confirm by fitting to scaling functions.
We next investigate how second-order DPTs can appear in one-dimensional free Brownian motion by choosing the observable, which essentially captures the localization transition of the trajectories.
We discuss and confirm that the DPTs predicted for high dimensions can also be found when considering many Brownian particles at lower dimensions.
\end{abstract}

\maketitle

\section{Introduction}

The study of stochastic dynamics has yielded profound insights into the behavior of diverse systems, from the intricate workings of molecular motors within cells~\cite{Seifert2012stochastic}  to the collective fluctuations of magnets and fluids near criticality~\cite{Hohenberg1977}.
Traditionally, the focus has been on characterizing typical fluctuations, which dominate the average or variance of physical observables.
However, atypical fluctuations play a crucial role in a wide range of phenomena, including the occurrence of rare events~\cite{giardina2011simulating}, the dynamics of glassy systems~\cite{garrahan2007dynamical}, and the stability of non-equilibrium states~\cite{Espigares2013}.
In particular, exponentially rare fluctuations of time-averaged observables have attracted significant attention, with large deviation theory providing a powerful framework for their analysis~\cite{Touchette2009}.

A pivotal concept arising from the study of atypical fluctuations is the dynamical phase transition (DPT)~\cite{Touchette2009,garrahan2007dynamical}, a qualitative change in the dynamical path that emerges upon varying the value of a time-integrated observable.
This concept, analogous to thermodynamic phase transitions, manifests as a singularity in the dynamical free energy, also known as the large deviation rate function.
The order of a DPT is determined by the nature of the singularity in the Legendre-Fenchel transform of the rate function, termed the scaled cumulant generating function (SCGF)~\cite{Touchette2009}, which serves as the dynamical counterpart of the thermodynamic potential in the canonical ensemble~\cite{Chetrite2013}.
While second-order DPTs and associated critical phenomena have been explored in detail within kinetic spin models like the Ising model with the Glauber dynamics~\cite{Jack2010}, first-order DPTs have been observed for example in kinetically constrained models~\cite{garrahan2007dynamical,garrahan2009first}, which are considered as simplified models for glasses.

Unlike thermodynamic phase transitions that are defined in the large-size limit, DPTs can arise even in single-particle systems by considering the long-time limit.
A prime example is the first-order DPT observed in one-dimensional Brownian motion with drift~\cite{Nyawo2017,Nyawo2018}, which exhibits a connection to the localization transition in non-Hermitian quantum models~\cite{Hatano1997}.
Intriguingly, this drifting Brownian motion exhibits phase separation-like behavior in the temporal domain: when examining the fraction of time spent in a given region, a typical time frame emerges where the particle remains close to its starting point before eventually diffusing away~\cite{Nyawo2017}.
This observation raises a compelling question: can such DPT and phase separation-like behavior occur in an even simpler setting, namely, Brownian motion without drift?
This question is particularly relevant as it allows us to explore the connection between DPT and other kinds of phase transitions, such as the adsorption transition of a single polymer and depinning transition, potentially revealing a deeper understanding of the underlying principles governing these phenomena.

The mathematical framework used to analyze these atypical fluctuations involves solving an eigenvalue problem for the SCGF, a process similar to solving the Schr\"{o}dinger equation in quantum mechanics.
This correspondence essentially connects the classical DPT to the quantum phase transition, which has been recently studied in the context of many-body physics, for example in an attempt to understand and extend active matter~\cite{tociu2019dissipation, Adachi2022, Takasan2024}.
For the simpler single-body settings, extensive research has been conducted for the eigenvalue problem of these Schr\"{o}dinger-like equations, revealing interesting dimensionality-dependent behavior, particularly in the case of shallow potentials~\cite{Simon1976, Blankenbecler1977, Kocher1977, Klaus1977, Klaus1980, Klaus1980_2, Bronzan1987, Yang1989, Brownstein2000}.
This motivates us to investigate how, even for single-body problems, dimensionality can affect the phase behavior of the rate function, particularly in the context of Brownian motion without drift. 

We here present two distinct types of DPTs observed for the single-particle Brownian motion in the absence of external drift forces.
Firstly, we demonstrate the emergence of a first-order DPT with temporal phase separation for a simple Brownian motion at high dimensions, by finding scaling relations in the occupation order parameter.
The critical dimension for this first-order DPT is $d=4$, which is distinct from $d=2$ where the change in the behavior of recursiveness in Brownian motion takes place.
We confirm by numerical examples that the same transition can be observed for $N$ particles at $d$ dimensions as far as $Nd > 4$.

As a second example, we uncover the appearance of a localization transition as second-order DPTs in one-dimensional Brownian motion when measuring the difference in occupations between two regions.
These second-order DPTs exhibit universal exponents in the rate function, independent of the specific observable, hinting at a connection to the adsorption transition of a long polymer.

In the rest of the manuscript, we first briefly explain the general formulation of the large deviation theory (Sec.~\ref{sec:formulation}).
Then, we present a model of Brownian motion that exhibits a first-order dynamical phase transition in its large deviations depending on spatial dimensions (Sec.~\ref{sec:dimension}).
We analytically derive asymptotic behaviors of the SCGF and rate functions for each dimension (Sec.~\ref{subsec:analytical}) and further confirm the predicted temporal phase separations in high dimensions through simulations (Sec.~\ref{sec_dim_simulation}).
Our results for spatial dimensions are then extended to the case of large deviations of several particles, where we find similar dimension-dependent behaviors can be observed with appropriate observables (Sec.~\ref{sec:number}).
Next, we consider a simple case with $d=1$, focusing on the behaviors of rate functions around typical conditions, identifying specific leading order behaviors and related singularities, and connecting these singularities to localization transitions (Sec.~\ref{sec:universality}).
Finally, we draw conclusions and provide related discussions (Sec.~\ref{sec:conclusion}).

\begin{figure}[t]
    \centering
    \includegraphics[scale=1]{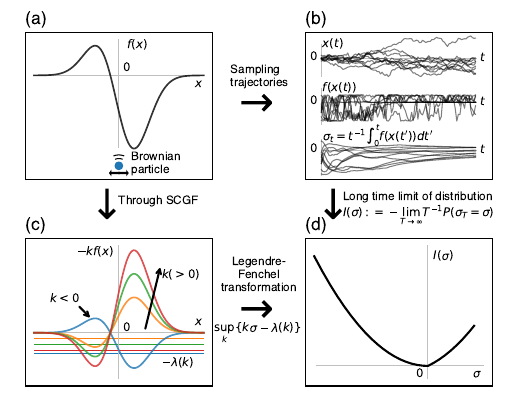}
    \caption{Schematic figures of the large deviation problem considered throughout the paper.
    (a) $f(x)$ is a scalar observable that we choose, which is a function of the stochastic variable $x$ undergoing Brownian motion.
    (b) $x(t)$, $f\bm{(}x(t)\bm{)}$, and $\sigma_t = t^{-1} \int_0^t f\bm{(}x(t')\bm{)} dt'$ are shown as a function of time $t$.
    We take $\sigma_T$ as a time-averaged observable and consider its distribution as $T \to \infty$, where $T$ is the total observation time.
    (c) The SCGF $\lambda(k)$ is obtained as the dominant eigenvalue of the problem $\mathcal{L}_k \phi_k(x) = [ \nabla^2 + kf(x) ] \phi_k(x) = \lambda(k) \phi_k(x)$.
    This eigenvalue problem is equivalent to the quantum eigenenergy problem for the ground state of a single particle in a potential $-kf(x)$ up to a minus sign, in which case the obtained eigenenergy is $-\lambda(k)$.
    We obtain $\lambda(k)$ as a function of $k$ upon changing $k$ from $-\infty$ to $\infty$.
    (d) The rate function $I(\sigma) := - \lim_{T \to \infty} T^{-1} \ln P(\sigma_T=\sigma)$, where $P(\sigma_T=\sigma)$ is the probability density at $\sigma_T = \sigma$.
    The rate function can also be obtained from the Legendre-Fenchel transform of the SCGF: $I(\sigma) = \sup_{k} \{ k \sigma - \lambda(k) \}$.}
    \label{fig:schematic}
\end{figure}

\section{Formulation}
\label{sec:formulation}

In this paper, we consider the Brownian motion of a particle in $d$ dimensions:
\begin{equation}
    \frac{d\bm{x}(t)}{dt} = \bm{\xi}(t),
    \label{eq:SDE}
\end{equation}
where $\bm{x}(t)$ represents the position of the particle, $\bm{\xi}(t)$ is a $d$-dimenional Gaussian noise with $\braket{\xi_a(t)} = 0$ and $\braket{\xi_a(t) \xi_b(t')} = 2 \delta_{ab}\delta(t-t')$ ($a, b \in \{ 1,2,\cdots,d \}$).
$\delta(\cdot)$ is the Dirac delta function.
Here, the diffusion constant is set to unity by rescaling time, and we set the initial condition as $\bm{x}(0) = \bm{0}$.
As a time-averaged observable, we consider
\begin{equation}
    \sigma_T := \frac{1}{T} \int^T_0 f\bm{(}\bm{x}(t)\bm{)} dt,
    \label{eq_observable}
\end{equation}
where $f(\bm{z})$ is a general scaler function of a $d$-dimensional variable $\bm{z}$, and $T$ is the total observation time [Figs.~\ref{fig:schematic}(a) and (b)].
If $f(\bm{z})$ approaches zero [$f(\bm{z}) \to 0$] sufficiently quickly as $||\bm{z}|| \to \infty$, $\sigma_T$ converges in probability to zero as $T \to \infty$.
For large but finite $T$, the probability density of $\sigma_T$, $P(\sigma_T=\sigma)$, follows
\begin{equation}
    P(\sigma_T=\sigma) = e^{-TI(\sigma) + o(T)},
\end{equation}
according to the large deviation principle~\cite{Touchette2009, Touchette2018}.
Here, the rate function $I(\sigma)$ is defined as (see Fig.~\ref{fig:schematic})
\begin{equation}
    I(\sigma) := -\lim_{T \to \infty} \frac{1}{T} \ln{P(\sigma_T=\sigma)}.
\end{equation}

We can obtain $I(\sigma)$ from the SCGF $\lambda(k)$:
\begin{equation}
    \lambda(k) := -\lim_{T \to \infty} \frac{1}{T} \ln{\langle e^{Tk\sigma_T} \rangle}.
\end{equation}
Here, $k \in \mathbb{R}$ is a conjugate variable to $\sigma$.
Indeed, if $I(\sigma)$ is convex, $I(\sigma)$ and $\lambda(k)$ are related by the Legendre-Fenchel transformation from the G\"{a}rtner-Ellis theorem~\cite{Touchette2009}:
\begin{equation}
    I(\sigma) = \sup_{k} \{k\sigma-\lambda(k)\}.
    \label{eq:L-F}
\end{equation}
If $\lambda(k)$ is convex, Eq.~\eqref{eq:L-F} can be written as
\begin{equation}
    I(\sigma) = k^*\sigma - \lambda(k^*),
    \label{eq:Legendre}
\end{equation}
where $k^*$ satisfies
\begin{equation}
    \lambda'(k^*-0) \leq \sigma \leq \lambda'(k^*+0)
    \label{eq:k_star}
\end{equation}
with $\lambda'(k) := d \lambda (k) / d k$.
When $I(\sigma)$ and $\lambda(k)$ are convex,
\begin{equation}
    \lambda(k) = \sup_{\sigma} \{\sigma k-I(\sigma)\}
    \label{eq:L-F_lambda}
\end{equation}
also holds [Fig.~\ref{fig:schematic}(d)].
Then, Eq.~\eqref{eq:L-F_lambda} can be written as
\begin{equation}
    \lambda(k) = \sigma^*k - I(\sigma^*),
\end{equation}
where $\sigma^*$ satisfies
\begin{equation}
    I'(\sigma^*-0) \leq k \leq I'(\sigma^*+0)
\end{equation}
with $I'(\sigma) := d I(\sigma) / d \sigma$.

The order of DPTs is defined as the order of singularity of the SCGF $\lambda (k)$.
For example, if the first derivative of $\lambda (k)$ is discontinuous, it is called the first-order DPT.
If $\lambda (k)$ is undifferentiable, $I(\sigma)$ is not necessarily convex in the region corresponding to the singularity in $\lambda (k)$.
In this case, the rate function calculated from the Legendre-Fenchel transform is a convex envelope of $I(\sigma)$; we need other methods such as simulations to confirm the convexity of $I(\sigma)$ itself.

For the $d$-dimensional Brownian motion [Eq.~\eqref{eq:SDE}] with the time-averaged observable [Eq.~\eqref{eq_observable}], we can obtain the SCGF $\lambda (k)$ as the dominant eigenvalue of the tilted generator:
\begin{equation}
    \mathcal{L}_k = \nabla ^2 + kf(\bm{x}),
\end{equation}
where $\nabla := \partial / \partial \bm{x}$.
The operator $\mathcal{L}_k$ and the eigenvalue $\lambda(k)$ correspond to the quantum Hamiltonian and ground-state energy with their signs reversed, respectively, of a particle in $d$-dimensional space subject to a potential of the form $-kf(\bm{x})$.
The dominant eigenfunction $\phi_k (\bm{x})$, which we take as a real function, satisfies
\begin{equation}
    \mathcal{L}_k \phi_k(\bm{x}) = \lambda(k) \phi_k(\bm{x}).
    \label{eq:spectrum}
\end{equation}
Note that the left and right eigenfunctions are identical up to a constant factor since $\mathcal{L}_k$ is Hermitian.

When $\lambda(k)$ is differentiable and the condition for the ensemble equivalence is satisfied~\cite{Chetrite2015,Agranov2020}, the probability density for $\bm{x}$ conditioned with $\sigma_T = \sigma$, $P(\bm{x} | \sigma_T = \sigma)$, satisfies
\begin{equation}
    P(\bm{x} | \sigma_T = \sigma) \propto \phi_{k^*(\sigma)}(\bm{x})^2,
    \label{eq_relation_dist_wavefunc}
\end{equation}
apart from the time domain close to $t = 0$ or $t = T$.
Here, $k^*(\sigma) := \mathop{\mathrm{argmax}}_k \, \{k\sigma - \lambda(k)\}$.
When $\phi_{k^*(\sigma)} (\bm{x})$ is a wavefunction localized at some region, the distribution $P(\bm{x} | \sigma_T = \sigma)$ should also be localized around the same region.

\begin{figure}[t]
    \centering
    \includegraphics[scale=1]{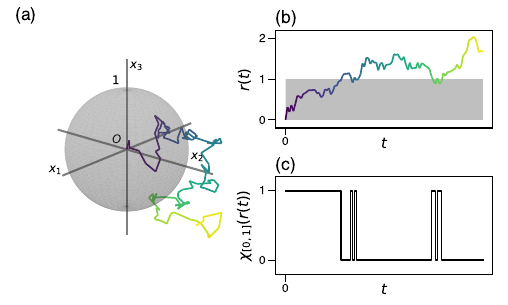}
    \caption{Schematic figures for the problem in Sec.~\ref{sec:dimension}.
    (a) Example path of a Brownian particle in three dimensions ($d = 3$).
    The particle starts from the origin as shown by the trajectory, where the brightness indicates time.
    The gray sphere suggests the region where $x_1^2 + x_2^2 + x_3^2 \leq 1$.
    (b) Time dependence of the distance from the origin, $r(t) = \sqrt{x_1(t)^2 + x_2(t)^2 + x_3(t)^2}$.
    The gray area corresponds to the gray sphere in (a).
    (c) Time dependence of $\chi_{[0,1]} \bm{(} r(t) \bm{)}$, which takes $1$ for $r(t) \leq 1$ and $0$ otherwise.
    The observable $\rho_T$ [Eq.~\eqref{eq:observable_rho}] is the time average of $\chi_{[0,1]} \bm{(} r(t) \bm{)}$, or the fraction of time spent by the particle inside the sphere.}
    \label{fig:schematic_3d}
\end{figure}

\begin{table*}[t]
\centering
\begin{tabular}{c || c |  c |  c |  c |  c |  c}
Dimension $d$ &  $1$  &  $2$ &  $3$  &  $4$  &  $5$ & $d$ ($\geq 5$) \\
\hline
SCGF $\lambda(k)$ for $\lambda(k) \approx 0$ &
$ k^2 $ & $ 4e^{-2\gamma}e^{-4/k} $ & $ \frac{1}{4} \left( k - \frac{\pi^2}{4} \right)^2 $ & $ - \frac{k - k_c^{(4)}}{\ln{\left( k - k_c^{(4)} \right)}} $ & $ \frac{1}{3} (k - \pi^2) $ & $ \frac{d-4}{d-2} (k - k_c^{(d)}) $ \\
\hline
Rate function $I(\rho)$ for $\rho \approx 0$ & $\approx \frac{1}{4} \rho^2$ & $\approx -4 \frac{\rho}{\ln{\rho}}$ & $\approx \frac{\pi^2}{4} \rho$ & $\approx k_c^{(4)} \rho$ & $=\pi^2 \rho$ $(0 < \rho < \frac{1}{3})$ & $= k_c^{(d)} \rho$ $(0 < \rho < \rho_c^{(d)} = \frac{d-4}{d-2})$ \\
\end{tabular}
\caption{Leading order behavior of the SCGF $\lambda(k)$ and the rate function $I(\rho)$ for different dimensions $d$.
When $d$ is five or higher, $\lambda(k)$ is undifferentiable at $k=k_c^{(d)}$, suggesting a first-order DPT; correspondingly, $I(\rho)$ is a strictly linear function for a finite domain $0 < \rho < \rho_c^{(d)}$, where the temporal phase separation is expected to appear.
As $d \to \infty$, we asymptotically obtain $k_c^{(d)} \approx d^2/4$ and $\rho_c^{(d)} \approx 1 - 2/d$.}
\label{tab_expanded_scgf}
\end{table*}

\section{Dimensionality-induced temporal phase separation}
\label{sec:dimension}

In this section, we consider the fraction of time spent by the Brownian particle within a $d$-dimensional ball centered at the origin (see Fig.~\ref{fig:schematic_3d}) as the observable $\sigma_T$ [Eq.~\eqref{eq_observable}]:
\begin{equation}
    \rho_T := \frac{1}{T} \int^T_0 \chi_{[0,1]}\bm{(}r(t)\bm{)} dt,
    \label{eq:observable_rho}
\end{equation}
where the radius of the ball is used as the length unit.
Here, $r(t) := ||\bm{x}(t)||$ with $r(0)=0$, and $\chi_A(z)$ is the indicator function defined as $\chi_A(z) = 1$ for $z \in A$ and $\chi_A(z) = 0$ otherwise.
The stochastic process for $r(t)$ is the $d$-dimensional Bessel process, which shows a qualitative change in the returning probability at $d = 2$; the process is transient for $d > 2$ but recurrent for $d \leq 2$~\cite{Borodin2002}.

\subsection{Analytical derivation of the SCGF and rate function}
\label{subsec:analytical}

The eigenvalue problem [Eq.~\eqref{eq:spectrum}] for the observable $\rho_T$ is~\cite{Angeletti2016}
\begin{equation}
    [\nabla ^2 + k\chi_{[0,1]}(||\bm{x}||)] \phi_k(\bm{x}) = \lambda(k) \phi_k(\bm{x}).
    \label{eq_eigen_rho}
\end{equation}
Since $\chi_{[0,1]}(||\bm{x}||)$ is spherically symmetric, Eq.~\eqref{eq_eigen_rho} is reduced to
\begin{equation}
    \left[ \frac{d^2}{dr^2} + \frac{d-1}{r} \frac{d}{dr} - \frac{l(l+d-2)}{r^2} + k\chi_{[0,1]}(r) \right] \phi_k(r) = \lambda(k) \phi_k(r)
\end{equation}
with $l=0,1,2,\dots$, or equivalently~\cite{Nieto2002},
\begin{equation}
    \left[ \frac{d^2}{dr^2} + F_k(r) \right] \Phi_k(r) = \lambda(k) \Phi_k(r),
\end{equation}
where
\begin{align}
    \Phi_k(r) &:= r^{(d-1)/2} \phi_k(r), \\
    F_k(r) &:= -\frac{\left( l+\frac{d-3}{2} \right) \left( l+\frac{d-1}{2} \right)}{r^2} + k\chi_{[0,1]}(r).
\end{align}
We used the fact that the dominant eigenfunction $\phi_k (\bm{x})$ depends only on $r = ||\bm{x}||$ [i.e., $\phi_k(\bm{x}) = \phi_k(r)$] due to the rotational symmetry of $\mathcal{L}_k$.
Note that increasing $d$ by $2$ is equivalent to increasing $l$ by $1$, according to the functional form of $F_k (r)$.

Setting $l = 0$ to focus on the dominant eigenvalue, we obtain
\begin{align}
    \left\{ \,
    \begin{aligned}
    &\left[ \frac{d}{dr^2}+\frac{d-1}{r} \frac{d}{dr} + \left( k-\lambda(k) \right) \right] \phi_{k}(r) = 0 & (r \leq 1) \\
    &\left[ \frac{d}{dr^2}+\frac{d-1}{r} \frac{d}{dr} - \lambda(k) \right] \phi_{k}(r) = 0 & (r > 1)
    \end{aligned}
    \right. .
    \label{eq:radial_eigen_eq}
\end{align}
The procedure to obtain the eigenvalue $\lambda (k)$ is similar to that in quantum mechanics.
By solving Eq.~\eqref{eq:radial_eigen_eq}~\cite{Nieto2002}, we obtain
\begin{align}
    \phi_k(r) &= 
    \left\{ \,
    \begin{aligned}
    &A_k \frac{J_{(d-2)/2}\bm{(}r \sqrt{k-\lambda(k)}\bm{)}}{[r \sqrt{k-\lambda(k)}]^{(d-2)/2}} & (r \leq 1) \\
    &B_k \frac{K_{(d-2)/2}\big{(}r \sqrt{\lambda(k)}\big{)}}{[r \sqrt{\lambda(k)}]^{(d-2)/2}} & (r>1)
    \end{aligned}
    \right.
\end{align}
with $J_\nu (z)$ and $K_\nu (z)$ being the Bessel function of the first kind and the modified Bessel function of the second kind, respectively, where $\nu$ is the order.
Here we used the conditions that $\phi_k(r)$ is finite with slope zero at $r = 0$ and normalizable.
The ratio of the coefficients $A_k/B_k$ and the eigenvalue $\lambda(k)$ are determined by the continuity conditions of $\phi_k(r)$ and $d\phi_k(r) / dr$ at $r = 1$:
\begin{equation}
    \frac{\sqrt{k-\lambda} J_{d/2}(\sqrt{k-\lambda})}{J_{(d-2)/2}(\sqrt{k-\lambda})} = \frac{\sqrt{\lambda} K_{d/2}(\sqrt{\lambda})}{K_{(d-2)/2}{(\sqrt{\lambda}})}.
    \label{eq:Bessel_boundary}
\end{equation}

If $\lambda (k) > 0$, the wavefunction $\phi_k (r)$ is a localized (i.e., bound) state around $r \in [0, 1]$, and the corresponding distribution $P(\bm{x} | \rho_T = \rho)$ is also localized in a similar region, according to Eq.~\eqref{eq_relation_dist_wavefunc}.
We define the threshold $k_c^{(d)}$, at which the localized state starts to appear [i.e., $\lambda (k) \to + 0$ for $k \to k_c^{(d)} + 0$].
For $d = 1$ or $d = 2$, we see $k_c^{(1)}=k_c^{(2)}=0$~\cite{Landau_quantum} from Eq.~\eqref{eq:Bessel_boundary}.
For $d \geq 3$, $k_c^{(d)}$ is a positive value~\cite{Sahu1989} and is determined by Eq.~\eqref{eq:Bessel_boundary} with $\lambda = 0$ and $k = k_c^{(d)}$:
\begin{equation}
    \sqrt{k_c^{(d)}}J_{d/2} \left(\sqrt{k_c^{(d)}} \right)-(d-2)J_{(d-2)/2} \left(\sqrt{k_c^{(d)}} \right) = 0.
    \label{eq:kc_eq}
\end{equation}

For $\lambda(k) \approx 0$, we can analytically obtain the asymptotic functional form of $\lambda(k)$, as shown in detail in the following subsections.
Here, we briefly summarize the results (see Table~\ref{tab_expanded_scgf}).
When $d = 1$ or $d = 2$, $\lambda (k)$ has no singularity.
When $d = 3$ or $d = 4$, $\lambda (k)$ smoothly changes from zero to positive at $k = k_c^{(d)}$ ($> 0$).
In contrast, when $d \geq 5$, $\lambda (k)$ changes from zero to positive as $\lambda (k) \propto k - k_c^{(d)}$ for $k \geq k_c^{(d)}$ ($> 0$), suggesting a first-order DPT at $k = k_c^{(d)}$, where the first derivative of $\lambda(k)$ is discontinuous.
As $d$ is increased, we asymptotically obtain $\lambda (k) \approx (1-2/d)(k-d^2/4)$.
Note that the obtained $d$ dependence of $\lambda (k)$ is consistent with the previous work on single particle quantum systems~\cite{Klaus1980}.

The DPT seen in $\lambda (k)$ can lead to a singularity in the rate function $I(\rho)$ through the Legendre-Fenchel transformation.
For $d \leq 4$, $\lambda (k)$ is differentiable, and thus $I(\rho)$ is a smooth function for any $\rho > 0$, where no qualitative change in the dynamical path is expected.
In contrast, for $d \geq 5$, the above-mentioned first-order DPT suggests that $I(\rho)$ is strictly linear [i.e., $I(\rho) \propto \rho$ without subleading order terms] for $0 < \rho < \rho_c^{(d)}$ with a threshold $\rho_c^{(d)}$.
From Eq.~\eqref{eq:Legendre}, $\rho_c^{(d)}$ is obtained as $\rho_c^{(d)} = d \lambda(k) / d k |_{k \to k_c^{(d)} + 0}$.
In Sec.~\ref{sec_dim_simulation}, we will explain the consequences of this first-order DPT in the dynamical path of Brownian motion.

According to the correspondence between the bound state formation of a quantum particle and the adsorption of a polymer~\cite{Birshtein1991}, we can use the $d$-dependent property of $\lambda (k)$ to predict a polymer property for general dimensions; a Gaussian chain in a potential proportional to $-\chi_{[0,1]}(||\bm{x}||)$ should undergo the second-order adsorption transition in three or four dimensions and the first-order adsorption transition in five or higher dimensions.

In the following, for each dimension, we show the detailed derivation of the SCGF $\lambda(k)$ for small $k-k_c^{(d)}$ ($\geq 0$) and the corresponding rate function $I(\rho)$ for small $\rho$ ($\geq 0$).
We focus on the case with $k \geq 0$ since $\lambda(k)=0$ for $k \leq 0$.

\subsubsection{Derivation for $d=1$}
\label{sec:derivation_1d}

When $\nu$ is a half-integer, $J_{\nu}(z)$ and $K_{\nu}(z)$ are given with trigonometric functions~\cite{Abramowitz1965}, and in particular,
\begin{align}
    J_{-1/2}(z) &= \sqrt{\frac{2z}{\pi}} \frac{\cos{z}}{z}, \\
    J_{1/2}(z) &= \sqrt{\frac{2z}{\pi}} \frac{\sin{z}}{z},
    \label{eq:J_1/2} \\
    K_{-1/2}(z) = K_{1/2}(z) &= \sqrt{\frac{2z}{\pi}} \left( \frac{\pi}{2z}\right) e^{-z}.
    \label{eq:K_1/2}
\end{align}
Applying these to Eq.~\eqref{eq:Bessel_boundary}, we obtain $k_c^{(1)}=0$ and $\sqrt{k-\lambda}\tan{(\sqrt{k-\lambda})} = \sqrt{\lambda}$, which leads to
\begin{equation}
    \lambda (k) \approx k^2 \ \ \ (k \ll 1).
\end{equation}

As $\lambda (k)$ is continuous and differentiable everywhere, there is no singularity in $\lambda (k)$ as well as its Legendre-Fenchel transform $I(\rho)$.
This suggests that no DPT is found when $d = 1$.
We can obtain the asymptotic functional form of the rate function by the Legendre-Fenchel transformation of $\lambda (k)$ for small $k$, which is straightforward using Eq.~\eqref{eq:Legendre} as $\lambda (k)$ is continuous.
From $\rho=\lambda'(k^*)$, we obtain $k^* (\rho) \approx \rho / 2$, leading to
\begin{equation}
    I(\rho) = k^*(\rho) \rho - \lambda\bm{(}k^*(\rho)\bm{)} \approx \frac{1}{4} \rho^2 \ \ \ (\rho \ll 1).
\end{equation}

\subsubsection{Derivation for $d=2$}

For general $\nu \in \mathbb{R}$, $J_{\nu}(z)$ and $K_{\nu}(z)$ can be expressed as~\cite{Abramowitz1965}:
\begin{align}
    J_{\nu} (z) &= \left( \frac{1}{2}z \right)^\nu \sum_{k=0}^{\infty} \frac{\left( -\frac{1}{4}z^2 \right)^k}{k! \Gamma (\nu+k+1)},
    \label{eq:J_nu} \\
    K_{\nu}(z) &= \frac{1}{2}\pi \frac{I_{-\nu}(z)-I_{\nu}(z)}{\sin{(\nu \pi)}} \ \ \ (\nu \neq 0, \pm1, \dots)
    \label{eq:K_nu}
\end{align}
with
\begin{equation}
    I_{\nu}(z) = \left( \frac{1}{2}z \right)^\nu \sum_{k=0}^{\infty} \frac{\left( \frac{1}{4}z^2 \right)^k}{k! \Gamma(\nu+k+1)},
    \label{eq:I_nu}
\end{equation}
and for $n \in \mathbb{Z}_{\geq 0}$,
\begin{align}
    K_n(z) &= \frac{1}{2} \left( \frac{1}{2}z \right)^{-n} \sum_{k=0}^{n-1} \frac{(n-k-1)!}{k!} \left( -\frac{1}{4}z^2 \right)^k +(-)^{n+1} \ln{\left( \frac{1}{2}z \right)} I_n(z) \nonumber \\
    & \quad + (-)^n \frac{1}{2} \left( \frac{1}{2}z \right)^n \sum_{k=0}^{\infty} \{ \psi(k+1)+\psi(n+k+1) \} \frac{\left( \frac{1}{4}z^2 \right)^k}{k! (n+k)!}
    \label{eq:K_n}
\end{align}
with the polygamma functions
\begin{align}
    \psi(1) &= -\gamma, \\
    \psi(n) &= -\gamma + \sum_{k=1}^{n-1} k^{-1} \ \ \ (n \geq 2),
\end{align}
where $\gamma$ is Euler's constant.
Here, $\Gamma (z)$ is the gamma function.

To solve Eq.~\eqref{eq:Bessel_boundary} for $d = 2$, we consider $\nu = (d-2)/2 = 0$ and $\nu = d/2 = 1$.
For $z \ll 1$, we obtain
\begin{align}
    J_0 (z) &\approx 1 - \frac{1}{4} z^2,
    \label{eq:J_0} \\
    J_1 (z) &\approx \frac{1}{2} z - \frac{1}{16} z^3,
    \label{eq:J_1} \\
    K_0 (z) &\approx - \left[ \ln{\left( \frac{1}{2}z \right)} + \gamma \right],
    \label{eq:K_0} \\
    K_1 (z) &\approx z^{-1} + \frac{1}{2} z \ln{\left( \frac{1}{2} z \right)}.
    \label{eq:K_1}
\end{align}
By substituting these into Eq.~\eqref{eq:Bessel_boundary}, we obtain $k_c^{(2)}=0$, and
\begin{equation}
    \lambda (k) \approx 4e^{-2\gamma}e^{-4/k} \ \ \ (k \ll 1).
    \label{eq_lambda_2d}
\end{equation}

We obtain the derivative of $\lambda(k)$ as $\lambda'(k) \approx 4C_0 k^{-2} e^{-4 / k}$ with $C_0 := 4 e^{-2\gamma}$, leading to
\begin{equation}
    k^*(\rho) \approx -\frac{4}{2W(-\sqrt{\rho/C_0})} \approx -\frac{4}{\ln{\rho} - 2\ln{(-\ln{\rho})}},
\end{equation}
where $W(z)$ is the Lambert W function on branch $-1$, usually denoted by $W_{-1}(z)$.
Here we used~\cite{Olver2010, Corless1996lambert}
\begin{equation}
    W(-z) \approx \ln z - \ln (-\ln z) \ \ \ (z \ll 1).
    \label{eq:Lambert}
\end{equation}
From $I(\rho) = k^*(\rho) \rho - \lambda\bm{(}k^*(\rho)\bm{)}$, we obtain
\begin{equation}
    I(\rho) \approx -\frac{4 \rho}{\ln{\rho} - 2\ln{(-\ln{\rho})}} \approx -\frac{4\rho}{\ln{\rho}} \ \ \ (\rho \ll 1).
\end{equation}
As $\lambda(k)$ is differentiable for $k > 0$, the corresponding $I(\rho)$ does not have a strictly linear region.
Also, it is clear from Eq.~\eqref{eq_lambda_2d} that $\lambda (k)$ shows no DPT.

\subsubsection{Derivation for $d=3$}

For $d=3$, we can use~\cite{Abramowitz1965}
\begin{equation}
    J_{3/2}(z) = \sqrt{\frac{2z}{\pi}} \left( \frac{\sin{z}}{z^2}-\frac{\cos{z}}{z} \right)
    \label{eq:J_3/2}
\end{equation}
and Eq.~\eqref{eq:J_1/2} for Eq.~\eqref{eq:kc_eq} to obtain
\begin{equation}
    \cot{\left( \sqrt{k_c^{(3)}} \right)} = 0,
\end{equation}
which leads to $k_c^{(3)}=\pi^2/4$.

Using Eqs.~\eqref{eq:J_1/2}, \eqref{eq:K_1/2}, \eqref{eq:J_3/2}, and~\cite{Abramowitz1965}
\begin{equation}
    K_{3/2}(z) = \sqrt{\frac{2z}{\pi}} \left( \frac{\pi}{2z}\right) e^{-z} (1+z^{-1})
    \label{eq:K_3/2}
\end{equation}
in Eq.~\eqref{eq:Bessel_boundary}, we obtain
\begin{equation}
    \sqrt{k-\lambda}\cot{(\sqrt{k-\lambda})} = -\sqrt{\lambda}.
    \label{eq:Bessel_boundary_3d}
\end{equation}
To solve this equation for $k \approx k_c^{(3)} = \pi^2 / 4$, we define $\delta := k-k_c^{(3)}$ $(\geq 0)$ and assume $\delta, \lambda \ll 1$.
Applying $\sqrt{k - \lambda} = \sqrt{\pi^2 / 4 + \delta - \lambda} = \pi / 2 + \delta / \pi + O(\delta^2, \lambda)$ to Eq.~\eqref{eq:Bessel_boundary_3d}, we obtain $-\delta / 2 + O(\delta^2, \lambda) = - \sqrt{\lambda}$, which leads to
\begin{equation}
    \lambda(k) \approx \frac{1}{4}(k - k_c^{(3)})^2 = \frac{1}{4} \left(k-\frac{\pi^2}{4} \right)^2 \ \ \ \left( 0 \leq k - \frac{\pi^2}{4} \ll 1 \right).
\end{equation}

From $\lambda'(k) \approx (k-k_c^{(3)})/2$, we obtain $k^*(\rho) = 2 \rho + k_c^{(3)}$ and
\begin{equation}
    I(\rho) \approx k_c^{(3)}\rho + \rho^2 = \frac{\pi^2}{4}\rho + \rho^2 \ \ \ (\rho \ll 1).
\end{equation}
Note that $\lambda(k)$ is differentiable and so is $I(\rho)$.
Also, $I(\rho)$ does not have strictly linear regions because of the differentiability of $\lambda (k)$.
However, the second derivative of $\lambda(k)$ is not continuous at $k = k_c^{(3)}$, which suggests a second-order DPT.
The second-order DPT is considered a transition from a non-localized state to a localized state by analogy with the corresponding quantum problem.

\subsubsection{Derivation for $d=4$}

For $d=4$, we can show for $\delta, \lambda \ll 1$,
\begin{equation}
    \frac{\sqrt{k-\lambda} J_2(\sqrt{k-\lambda})}{J_1(\sqrt{k-\lambda})} = 2 + \frac{1}{2} (\delta - \lambda) + O(\delta^2, \lambda^2),
    \label{eq:4d_lhs}
\end{equation}
using Eqs.~\eqref{eq:Bessel_boundary} and \eqref{eq:kc_eq} (see Sec.~\ref{sec_dim_derivation_6d} for the derivation), where $\delta := k-k_c^{(4)}$ with $k_c^{(4)}$ ($\approx 5.78$ from numerical calculation) determined from Eq.~\eqref{eq:kc_eq}.
We use Eq.~\eqref{eq:K_1} and
\begin{equation}
    K_2(z) \approx 2z^{-2}\left(1-\frac{1}{4}z^2\right)
    \label{eq:K_2}
\end{equation}
for $z \ll 1$ to obtain
\begin{equation}
    \frac{\sqrt{\lambda} K_2(\sqrt{\lambda})}{K_1(\sqrt{\lambda})} \approx 2 - \frac{1}{2} \lambda \ln{\lambda}.
    \label{eq:4d_rhs}
\end{equation}
Thus, from Eqs.~\eqref{eq:4d_lhs} and \eqref{eq:4d_rhs} with Eq.~\eqref{eq:Bessel_boundary}, we get
\begin{equation}
    \delta \approx -\lambda \ln{\lambda}.
\end{equation}
To obtain $\lambda$ as a function of $\delta$ (or $k$), we use the Lambert W function [Eq.~\eqref{eq:Lambert}]:
\begin{align}
    \lambda (k) &\approx -\frac{\delta}{W(-\delta)} \approx -\frac{\delta}{\ln{\delta} - \ln{(-\ln{\delta})}} \approx - \frac{\delta}{\ln{\delta}} \\
    &= - \frac{k-k_c^{(4)}}{\ln{(k-k_c^{(4)})}} \ \ \ (0 \leq k-k_c^{(4)} \ll 1).
\end{align}

For $d = 4$, as well as the case of $d = 2$, the SCGF $\lambda(k)$ is not expressed as a power function.
Therefore, for $\lambda(k) \approx -(k-k_c^{(4)})/\ln{(k-k_c^{(4)})}$, we get 
\begin{equation}
    \lambda'(k) \approx -\frac{1}{\ln{(k-k_c^{(4)})}} + \frac{1}{[\ln{(k-k_c^{(4)})}]^2}.
\end{equation}
By solving $\rho=\lambda'(k^*)$, we get
\begin{equation}
    \ln{(k^*-k_c^{(4)})} \approx  \frac{-1 \pm \sqrt{1+4\rho}}{2\rho}.
\end{equation}
Since $k^*-k_c^{(4)} \ll 1$ and thus $\ln{(k^*-k_c^{(4)})}<0$, we choose the minus sign as $\ln{(k^*-k_c^{(4)})} \approx - ( 1 + \sqrt{1 + 4 \rho} )/(2\rho)$, leading to
\begin{equation}
    k^*-k_c^{(4)} \approx \exp{\left( - \frac{ 1 + \sqrt{1 + 4\rho}}{2\rho} \right)}.
\end{equation}
Then $I(\rho)$ is given as
\begin{equation}
    I(\rho) \approx k_c^{(4)} \rho + e^{-1/\rho + o(1/\rho)} \ \ \ (\rho \ll 1),
\end{equation}
which does not have strictly linear regions.

\subsubsection{Derivation for $d \geq 5$}
\label{sec_dim_derivation_6d}

To consider the general asymptotic form of $\lambda (k)$ and $I(\rho)$ for $d \geq 5$, we assume $0 \leq \delta := k - k_c^{(d)} \ll 1$ and $\lambda \ll 1$.
Let us evaluate the left-hand side of Eq.~\eqref{eq:Bessel_boundary}.
As $\sqrt{k-\lambda} = \sqrt{k_c^{(d)} + \delta - \lambda } = \sqrt{k_c^{(d)}} + \left. (\delta - \lambda ) \middle/ \left( 2\sqrt{k_c^{(d)}} \right) \right. + O(\delta^2, \lambda^2)$, we get
\begin{align}
    & \quad J_\nu \left( \sqrt{k_c^{(d)} + \delta - \lambda} \right) \nonumber \\
    &= J_\nu \left( \sqrt{k_c^{(d)}} \right) + (\delta - \lambda) \left. J'_\nu \left( \sqrt{k_c^{(d)}} \right) \middle/ \left( 2\sqrt{k_c^{(d)}} \right) \right. + O (\delta^2, \lambda^2).
    \label{eq:J_nu_approx}
\end{align}
Thus, the left-hand side of Eq.~\eqref{eq:Bessel_boundary} is reduced to
\begin{align}
    &\quad \left. \sqrt{k_c^{(d)} + \delta - \lambda} \ J_{d/2} \left( \sqrt{k_c^{(d)} + \delta - \lambda} \right) \middle/ J_{(d-2)/2} \left( \sqrt{k_c^{(d)} + \delta - \lambda} \right) \right. \nonumber \\
    &= (d-2) + \frac{1}{2} (\delta - \lambda) + O(\delta^2, \lambda^2),
    \label{eq_boundary_lhs}
\end{align}
where we used Eqs.~\eqref{eq:kc_eq} and \eqref{eq:J_nu_approx} and the following relations~\cite{Abramowitz1965}:
\begin{align}
    J'_\nu (z) &= J_{\nu-1} (z) - \frac{\nu}{z} J_\nu (z) \label{eq_Jp_rel}, \\
    J'_\nu (z) &= -J_{\nu+1} (z) + \frac{\nu}{z} J_\nu (z).
\end{align}
Note that the above result also applies to the cases of $d = 3$ and $d = 4$, where $k_c^{(d)} > 0$.

The right-hand side of Eq.~\eqref{eq:Bessel_boundary} can be evaluated as follows.
For odd $d$ ($=5,7,9,\dots$), $d/2$ and $(d-2)/2$ are half-integers, and thus $\sin(d\pi/2)=-\sin\bm{(}(d-2)\pi/2\bm{)}=\pm 1$ for $d \equiv \pm 1 \pmod 4$.
Thus, for $\lambda \ll 1$, using Eqs.~\eqref{eq:K_nu} and \eqref{eq:I_nu}, the right-hand side of Eq.~\eqref{eq:Bessel_boundary} becomes 
\begin{align}
    &\quad \frac{\sqrt{\lambda} K_{d/2}(\sqrt{\lambda})}{K_{(d-2)/2}{(\sqrt{\lambda}})} \nonumber \\
    &= -\sqrt{\lambda} \frac{I_{-d/2}(\sqrt{\lambda})-I_{d/2}(\sqrt{\lambda})}{I_{-(d-2)/2}(\sqrt{\lambda})-I_{(d-2)/2}(\sqrt{\lambda})} \\
    &= - \sqrt{\lambda} \frac{ \left(\frac{1}{2}\sqrt{\lambda} \right)^{-d/2} \left[ \frac{1}{\Gamma (-d/2+1)} + \frac{1}{4 \Gamma (-d/2+2)} \lambda + O(\lambda^2) \right] }{\left(\frac{1}{2}\sqrt{\lambda} \right)^{-(d-2)/2} \left[ \frac{1}{\Gamma (-(d-2)/2+1)} + \frac{1}{4 \Gamma (-(d-2)/2+2)} \lambda + O(\lambda^2) \right]} \\
    &= (d-2) + \frac{1}{d-4} \lambda + O (\lambda^2).
    \label{eq_boundary_rhs}
\end{align}
On the other hand, for even $d$ ($=6,8,10,\dots$), both $d/2$ and $(d-2)/2$ are integers.
We instead use Eq.~\eqref{eq:K_n}, which leads to the same behavior of the right-hand side of Eq.~\eqref{eq:Bessel_boundary} as the case of odd $d$ [Eq.~\eqref{eq_boundary_rhs}]:
\begin{align}
    &\quad\frac{\sqrt{\lambda} K_{d/2}(\sqrt{\lambda})}{K_{(d-2)/2}{(\sqrt{\lambda}})} \nonumber \\
    &= \sqrt{\lambda} \frac{\frac{1}{2} \left( \frac{1}{2} \sqrt{\lambda} \right)^{-d/2} \left[ (d/2-1)! - \frac{(d/2-2)!}{4} \lambda + O (\lambda^2) \right]}{\frac{1}{2} \left( \frac{1}{2} \sqrt{\lambda} \right)^{-(d-2)/2} \left[ (d/2-2)! - \frac{(d/2-3)!}{4} \lambda + o(\lambda) \right]} \\
    &= (d-2) + \frac{1}{d-4} \lambda + o(\lambda).
\end{align}
Therefore, for the lowest order of $\delta$, we get
\begin{equation}
    \lambda (k) \approx \frac{d-4}{d-2} \delta = \frac{d-4}{d-2} (k-k_c^{(d)}) \ \ \ (0 \leq k-k_c^{(d)} \ll 1),
    \label{eq_expanded_form_general}
\end{equation}
and $\lambda(k) = 0$ for $k < k_c^{(d)}$.

The undifferentiable point in  $\lambda(k)$ at $k = k_c^{(d)}$ leads to the corresponding $I(\rho)$ having a linear region.
This is because $\lambda'(k)$ jumps from $\lambda'(k_c^{(d)}-0) = 0$ to $\lambda'(k_c^{(d)}+0) = (d-4)/(d-2)$ which is nonzero, and for the region $0 < \rho < \rho_c^{(d)} := (d-4)/(d-2)$ it is $\lambda\bm{(}k^*(\rho)\bm{)}=\lambda(k_c^{(d)})=0$, from which it follows
\begin{equation}
    I(\rho) = k^*(\rho) \rho = k_c^{(d)} \rho \ \ \ \left(0 < \rho < \rho_c^{(d)} = \frac{d-4}{d-2} \right)
    \label{eq:lever}
\end{equation}
for this region of $\rho$; for $\rho_c^{(d)} < \rho < 1$, $I(\rho)$ is a convex function.
In the case of $d=3$, on the other hand, the leading order of $I(\rho)$ is linear for small $\rho$ but higher order terms exist.

\subsubsection{Asymptotic form for $d=5$}
\label{sec_dim_derivation_5d}

We here explicitly calculate the asymptotic form for $d=5$, which is the smallest dimension in which the first-order DPT appears.
The (modified) Bessel functions of order $5/2$ are given as~\cite{Abramowitz1965}
\begin{align}
    J_{5/2}(z) &= \sqrt{\frac{2z}{\pi}} \left[ \left( \frac{3}{z^3} - \frac{1}{z} \right) \sin{z} -\frac{3}{z^2} \cos{z} \right], \label{eq:J_5/2} \\
    K_{5/2}(z) &= \sqrt{\frac{2z}{\pi}} \left( \frac{\pi}{2z}\right) e^{-z} (1+3z^{-1}+3z^{-2}). \label{eq:K_5/2}
\end{align}
Using Eq.~\eqref{eq:J_3/2} and \eqref{eq:J_5/2} in Eq.~\eqref{eq:kc_eq}, we obtain
\begin{equation}
    \sin{\left( \sqrt{k_c^{(5)}} \right)} = 0,
\end{equation}
which leads to $k_c^{(5)}=\pi^2$.
Using Eqs.~\eqref{eq:J_3/2}, \eqref{eq:K_3/2}, \eqref{eq:J_5/2}, and \eqref{eq:K_5/2}, we rewrite the boundary condition [Eq.~\eqref{eq:Bessel_boundary}] as
\begin{align}
    \frac{(3-k+\lambda)\sin{(\sqrt{k-\lambda})-3\sqrt{k-\lambda}}\cos{(\sqrt{k-\lambda})}}{\sin{(\sqrt{k-\lambda})-\sqrt{k-\lambda}}\cos{(\sqrt{k-\lambda})}} \nonumber \\
    = \frac{\lambda+3\sqrt{\lambda}+3}{\sqrt{\lambda}+1}.
\end{align}
By expanding both sides of this equation in terms of $k - k_c^{(5)}$ and $\lambda$, we finally obtain
\begin{equation}
    \lambda (k) \approx \rho_c^{(5)} (k-k_c^{(5)}) = \frac{1}{3} (k-\pi^2) \ \ \ (0 \leq k - \pi^2 \ll 1).
\end{equation}
Substituting $k_c^{(5)}=\pi^2$ and $\rho_c^{(5)}=1/3$ into Eq.~\eqref{eq:lever} yields
\begin{equation}
    I(\rho) = \pi^2 \rho \ \ \ \left(0 < \rho < \frac{1}{3} \right).
\end{equation} 

\begin{figure}[t]
    \centering
    \includegraphics[scale=1]{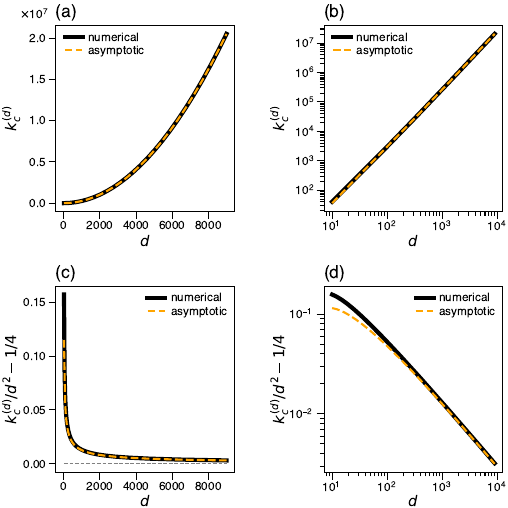}
    \caption{Dimension dependence of the threshold $k_c^{(d)}$.
    (a) Numerically obtained $d$ dependence of $k_c^{(d)}$ (black solid line), shown with the asymptotic form [Eq.~\eqref{eq_kc_largedim}] (orange dashed line).
    (b) Log-log plot of (a).
    (c) Numerically obtained $d$ dependence of $k_c^{(d)} / d^2 - 1 / 4$ (black solid line), shown with the asymptotic form (orange dashed line).
    (d) Log-log plot of (c).}
    \label{fig_highdim_asymp}
\end{figure}

\subsubsection{Asymptotic form for large $d$}
\label{sec_dim_asympt_law}

To check the $d$ dependence of $k_c^{(d)}$, we numerically solve Eq.~\eqref{eq:kc_eq} for $10 \leq d \leq 9000$ (black solid lines in Fig.~\ref{fig_highdim_asymp}), which suggests $k_c^{(d)} / d^2 \to 1/4 + 0$ ($d \to \infty$).
Setting $z_\nu := 2 \sqrt{k_c^{(d)}} / d $ and $\nu := d / 2$ in this subsection, we seek the asymptotic form of $z_\nu$ for $\nu \to \infty$ (i.e., asymptotic form of $k_c^{(d)}$ for $d \to \infty$).

We rewrite Eq.~\eqref{eq:kc_eq} as
\begin{equation}
    \frac{1}{z_\nu} + \frac{J'_\nu (\nu z_\nu)}{J_\nu (\nu z_\nu)} = \frac{\nu z_\nu}{2 \nu - 2},
    \label{eq_znu}
\end{equation}
where we use Eq.~\eqref{eq_Jp_rel}.
To consider $\nu \to \infty$, we use the asymptotic expansion of the Bessel function~\cite{Olver2010}:
\begin{align}
    & J_\nu (\nu z_\nu) \approx \left[ \frac{4 \zeta (z_\nu)}{{z_\nu}^2 - 1} \right]^{1/4} \frac{\mathrm{Ai} \bm{(}-\nu^{2 / 3} \zeta (z_\nu)\bm{)}}{\nu^{1/3}} \ \ \ (\nu \gg 1),
    \label{eq_j_asymp} \\
    & J'_\nu (\nu z_\nu) \approx -\frac{2}{z_\nu} \left[ \frac{{z_\nu}^2 - 1}{4 \zeta (z_\nu)} \right]^{1/4} \frac{\mathrm{Ai}' \bm{(}-\nu^{2 / 3} \zeta (z_\nu)\bm{)}}{\nu^{2/3}} \ \ \ (\nu \gg 1),
    \label{eq_jp_asymp}
\end{align}
where $\zeta (z) := [(3 / 2) \int_1^z dt (\sqrt{t^2 - 1} / t)]^{2 / 3}$, and $\mathrm{Ai}(z)$ is the Airy function.

We assume $z_\nu = 1 + \alpha \nu^{-2 / 3} + \beta \nu^{-1} + O (\nu^{-4 / 3})$ for $\nu \gg 1$ as an ansatz and determine the constants $\alpha$ and $\beta$ to satisfy Eq.~\eqref{eq_znu}.
Applying this ansatz to Eqs.~\eqref{eq_j_asymp} and \eqref{eq_jp_asymp} and noticing $\zeta (z_\nu) = 2^{1 / 3} \alpha \nu^{-2 / 3} + 2^{1 / 3} \beta \nu^{-1} + O (\nu^{-4 / 3})$, we obtain
\begin{equation}
    \frac{J'_\nu (\nu z_\nu)}{J_\nu (\nu z_\nu)} \approx -2^{1 / 3} \nu^{-1 / 3} \frac{\mathrm{Ai}' (-2^{1 / 3} \alpha)}{\mathrm{Ai} (-2^{1 / 3} \alpha) - 2^{1 / 3} \beta \mathrm{Ai}' (-2^{1 / 3} \alpha) \nu^{-1 / 3}}.
    \label{eq_j_ratio_expand}
\end{equation}
If $\mathrm{Ai}(-2^{1 / 3} \alpha) \neq 0$, Eq.~\eqref{eq_znu} would lead to $1 = 1 / 2$ for $\nu \to \infty$, which suggests that $-2^{1 / 3} \alpha$ should be one of zeros of the Airy function $\mathrm{Ai} (z)$.
Since we consider the dominant eigenvalue problem, $k_c^{(d)}$ (or $z_\nu$) should be the smallest when multiple candidates satisfy Eq.~\eqref{eq:kc_eq} [or Eq.~\eqref{eq_znu}].
Thus, $-2^{1 / 3} \alpha$ should be the first zero of $\mathrm{Ai} (z)$, written as $-a_1$ ($= -2.3381...$~\cite{Olver2010}).
Using $\alpha = 2^{-1 / 3} a_1$, we obtain $J'_\nu (\nu z_\nu) / J_\nu (\nu z_\nu) \to \beta^{-1}$ ($\nu \to \infty$) from Eq.~\eqref{eq_j_ratio_expand}.
Combining this with Eq.~\eqref{eq_znu} in the limit of $\nu \to \infty$, we obtain $1 + \beta^{-1} = 1 / 2$, leading to $\beta = -2$.

The obtained asymptotic form, $z_\nu \approx 1 + 2^{-1 / 3} a_1 \nu^{-2 / 3} - 2 \nu^{-1}$ ($\nu \gg 1$), is equivalent to
\begin{equation}
    \frac{k_c^{(d)}}{d^2} \approx \frac{1}{4} + \frac{a_1}{2^{2 / 3}} d^{-2 / 3} - 2 d^{-1} \ \ \ (d \gg 1).
    \label{eq_kc_largedim}
\end{equation}
We plot this result in Fig.~\ref{fig_highdim_asymp} with orange dashed lines, which agrees with the numerical result (black solid lines) for large $d$.

From Eqs.~\eqref{eq:lever} and \eqref{eq_kc_largedim}, we obtain the asymptotic form of $I (\rho)$ for large $d$:
\begin{equation}
    I (\rho) \approx \left[ \frac{d^2}{4} + O (d^{4 / 3}) \right] \rho \ \ \ \left( 0 \leq \rho \leq 1 - 2 d^{-1} + O (d^{-2}), \ d \gg 1 \right),
    \label{eq_I_highdim}
\end{equation}
which means that, as $d$ increases, the linear region for $I (\rho)$ expands and the slope of $I (\rho)$ increases.

\begin{figure*}[t]
    \centering
    \includegraphics[scale=1]{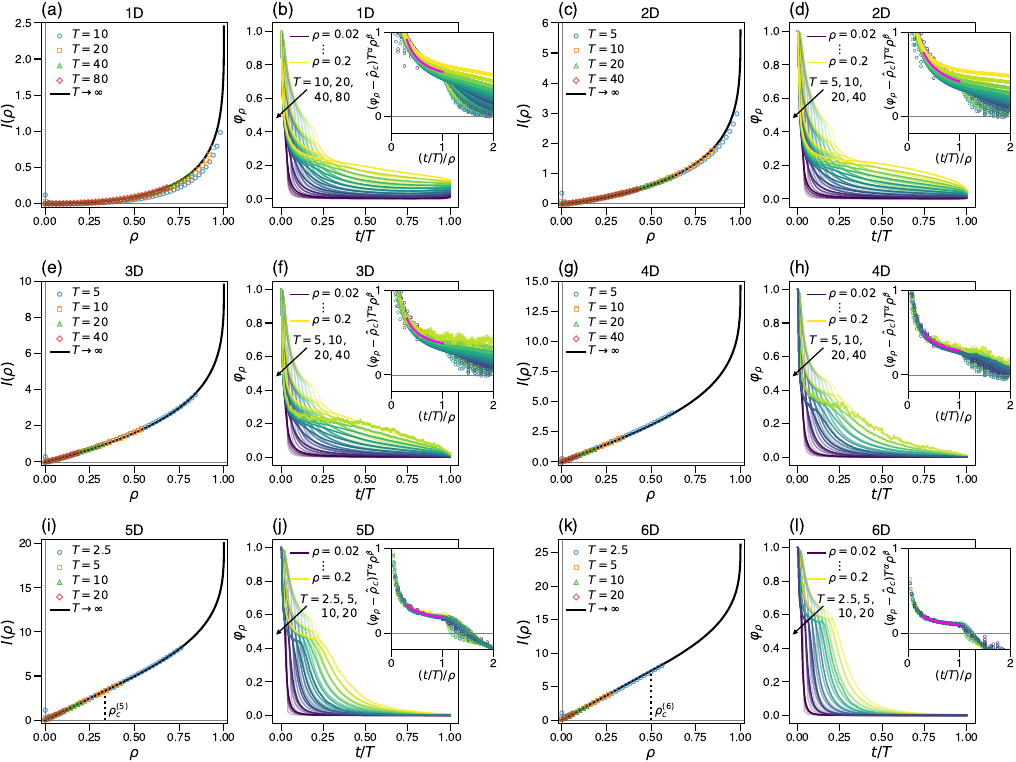}
    \caption{Dimensionality-induced phase separation.
    (a, c, e, g, i, k) Rate function $I(\rho)$ for $1 \leq d \leq 6$.
    In each panel, the colored symbols represent the data from simulations with several values of observation time $T$, and the solid line is the theoretical prediction for $T \to \infty$.
    In (i) $d = 5$ and (k) $d = 6$, the predicted $I(\rho)$ is strictly linear for $0 < \rho < \rho_c^{(d)}$, where phase separation is expected to appear; in contrast, no strictly linear region exists in lower dimensions.
    (b, d, f, h, j, l) Time dependence of the order parameter $\varphi_\rho (t)$ for each dimension.
    We plot the data obtained from simulations with different values of $T$, where the brightness suggests the size of $\rho$.
    The inset in each panel is a scaling plot of $\varphi_\rho (t)$ using the two exponents $\alpha > 0$ and $\beta > 0$ and the nominal critical point $\hat{\rho}_c$, which are obtained by fitting within the plateau-like region $t/T \in [0.3\rho, \rho]$ (magenta line suggesting the fitting curve).
    We set the time step as $dt = 0.05$ and took $10^{12}$ independent samples for all simulations, and the bin size for $\rho$ was set to $0.02$.}
    \label{fig_dimensionality_ps}
\end{figure*}

\subsection{Temporal phase separation in high dimensions}
\label{sec_dim_simulation}

It has been previously indicated~\cite{Nyawo2017,Nyawo2018} that the existence of a linear part in $I(\rho)$ leads to temporal phase separation (or dynamical phase coexistence) of the dynamical path: a particle trajectory separates into two segments, the first localized around the origin, and the second being non-localized.
This phenomenon is similar to the spatial phase separation in thermodynamics, where the Helmholtz free energy linearly depends on the particle density~\cite{Landau_statphys,jack2020ergodicity}.

The linear region with $I(\rho) = k_c^{(d)} \rho$ obtained for $d \geq 5$ indicates the temporal phase separation since the lever rule~\cite{Rubinstein2003,Nyawo2018} holds for $0 < p := \rho / \rho_c^{(d)} < 1$:
\begin{equation}
    I(\rho) = I(p\rho_c^{(d)}) = pI(\rho_c^{(d)}) + (1-p)I(0),
\end{equation}
where $p$ and $1 - p$ are the fractions of two phases with $\rho=\rho_c^{(d)}$ and $\rho = 0$, respectively.
The phase experiencing $\rho=\rho_c^{(d)}$ corresponds to a localized state (or a bound state in the quantum problem), where the particle stays close to the ball [see Fig.~\ref{fig:schematic_3d}(a)].
This phase represents atypical trajectories since such localization is very rare unless we condition the value of $\rho$.
On the other hand, the phase experiencing $\rho = 0$, where the particle is always outside of the ball, corresponds to a non-localized state (or an unbound state in the quantum problem) and represents typical trajectories.

To study whether the temporal phase separation appears in our setup with high enough dimensions, we performed simulations of Brownian motion for $1 \leq d \leq 6$.
In the left panels for each dimension of Fig.~\ref{fig_dimensionality_ps}, we plot $I(\rho)$ obtained from simulations (colored symbols), which agree with the theoretical predictions (black solid lines).
In particular, for $d = 5$ and $d = 6$ with $0 < \rho < \rho_c^{(d)} = (d - 4) / (d - 2)$ (see Sec.~\ref{sec_dim_derivation_6d}), the predicted linear dependence of $I(\rho)$ is asymptotically reproduced by the numerical results as $T$ increases.

To detect the phase separation, we define a time-dependent order parameter:
\begin{equation}
    \varphi_\rho (t) := \braket{\chi_{[0,1]}\bm{(}r(t)\bm{)}}_{\rho},
    \label{eq_order_param}
\end{equation}
which probes the localization in the $d$-dimensional unit ball centered at the origin, conditioned with $\rho_T = \rho$.
Here, $\braket{\cdots}_\rho$ denotes the expectation value with respect to the conditional probability $P(\bm{x}|\rho_T=\rho)$.
When the phase separation appears as $T \to \infty$ for $d \geq 5$, we expect temporally separated two phases: the localized phase with $\varphi_\rho(t) = \rho_c^{(d)}$ for the time domain $t/T \in (0, \rho/\rho_c^{(d)})$ and the non-localized phase with $\varphi_\rho(t) = 0$ for the remaining time domain $t/T \in (\rho/\rho_c^{(d)}, 1)$.
As a key feature of phase separation, $\varphi_\rho (t)$ in each phase does not depend on the specific value of $t/T$ or $\rho$, as long as $T$ is large enough.
The initial condition [$r(0) = 0$] sets the localized phase to appear in the first time domain [i.e., $t/T \in (0, \rho/\rho_c^{(d)})$].
Note that $T^{-1} \int_0^T dt \varphi_\rho(t) = \rho$ for any $\rho \in (0, \rho_c^{(d)})$, which is consistent with the condition $\rho_T = \rho$.
For $d \gg 1$, as $\rho_c^{(d)} \approx 1$ [see Eq.~\eqref{eq_I_highdim}], we expect phase separation between two phases with $\varphi_{\rho}(t) \approx 1$ and $\varphi_{\rho}(t)=0$.

In the right panels for each dimension of Fig.~\ref{fig_dimensionality_ps}, we plot the time dependence of $\varphi_\rho(t)$ with several values of $\rho$ and $T$.
For $d = 5$ and $d = 6$, $\varphi_\rho(t)$ shows a plateau-like behavior with a $\rho$-independent height as $T$ increases.
According to the property of phase separation, the height of this plateau-like region should approach $\rho_c^{(5)}$ ($= 1/3$) or $\rho_c^{(6)}$ ($= 1/2$) as $T \to \infty$, independently of $t/T$ or $\rho$.

To confirm the predicted behavior in the plateau-like region, we looked for a scaling law for $T \to \infty$ by fitting $\varphi_\rho (t)$.
Assuming a power-law scaling for $T$ and $\rho$, we fitted the observed $\varphi_\rho (t)$ by a function $F(t, T, \rho) := \hat{\rho}_c + T^{-\alpha} \rho^{-\beta} f\bm{(}(t/T)/\rho\bm{)}$ with $f(x) := (a_0 + a_1 x) / (1 + b_1 x)$, where $\alpha$, $\beta$, $\hat{\rho}_c$, $a_0$, $a_1$, and $b_1$ are fitting parameters.
Here, $\hat{\rho}_c$ denotes a nominal critical point, which is defined even for $d \leq 4$ and should be around the exact value $\rho_c^{(d)}$ for $d \geq 5$.
We assumed $\alpha, \beta > 0$ and $\hat{\rho}_c \in (0, 1)$, and chose $t/T \in [0.3\rho, \rho]$ for the plateau-like region used in fitting.
Using a higher-order rational function for $f(x)$ did not change the optimal values of $\alpha$, $\beta$, and $\hat{\rho}_c$ within the estimation error.
We used a Python package (scipy.optimize.curve\_fit~\cite{Scipy}) for fitting.

The scaling plots with the optimal fitting parameters for each dimension are shown in the insets of the right panels of Fig.~\ref{fig_dimensionality_ps}.
The magenta lines represent the best-fitted functions.
From Figs.~\ref{fig_dimensionality_ps}(j) and (l), we find clear scalings around the plateau-like region for $d = 5$ [with $(\alpha, \beta, \hat{\rho}_c) \approx (0.50, 0.46, 0.32)$] and $d = 6$ [with $(\alpha, \beta, \hat{\rho}_c) \approx (0.59, 0.57, 0.47)$], respectively.
This suggests that $\varphi_\rho - \hat{\rho}_c \sim T^{-\alpha} \to 0$, i.e., $\varphi_\rho \to \hat{\rho}_c$ ($\approx \rho_c^{(d)}$) for $T \to \infty$, consistent with the predicted phase separation for $d \geq 5$.
In contrast, we do not see a similar clear scaling for $d \leq 3$ [Figs.~\ref{fig_dimensionality_ps}(b), (d), and (f)], suggesting that there is no phase separation as expected.
For $d = 4$ [Fig.~\ref{fig_dimensionality_ps}(h), with $(\alpha, \beta, \hat{\rho}_c) \approx (0.50, 0.43, 0.18)$], which is the lower critical dimension for temporal phase separation, we see some violation of the scaling, though longer time simulations are needed to confirm this.

The emergence of the first-order DPTs in this section is closely related to space dimensionality.
Intuitively, the harder a particle returns to the original position, the more we observe the lower-order DPTs.
However, recursiveness of Brownian motion is not considered to be directly related.
Brownian motion is recursive when $d=1, 2$, while not for $d \geq 3$.
The first-order DPTs require more severe conditions to prevent the particle from returning to its original position since we find the lower critical dimension is $d=4$.

The same type of $d$-dependent DPT is expected to appear not only for the simple observable considered so far but also for general observables as long as the integrand $f(\bm{x})$ in Eq.~\eqref{eq_observable} vanishes sufficiently quickly as $||\bm{x}|| \to \infty$, according to the solution for the corresponding quantum problem~\cite{Klaus1980}.
Furthermore, we do not need to restrict ourselves to large deviations of \emph{one} Brownian particle.
In the next section, we show that \emph{several} low-dimensional Brownian particles without interactions are equivalent to a high-dimensional single particle if we take an appropriate observable.

\begin{figure}[t]
    \centering
    \includegraphics[scale=1]{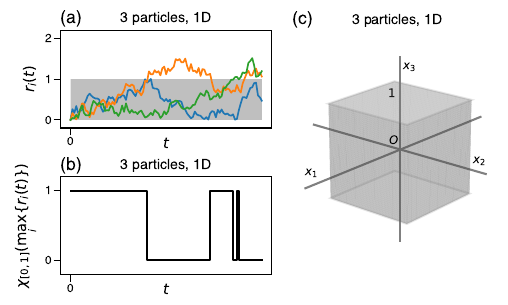}
    \caption{Schematic figures for the problem in the multi-particle setup.
    The illustrations are specific to $N = 3$ (total particle number) and $d=1$ (dimension).
    (a) Typical paths of the independent Brownian particles, $\{ r_i (t) \}_{i=1}^N$, where $r_i (t) = |x_i (t)|$ in one dimension.
    (b) Time dependence of the integrand of the observable, $\chi_{[0, 1]}\bm{(} \max_i \{ r_i(t) \} \bm{)}$, which takes $1$ for $r_i \leq 1$ ($\forall i$) and $0$ otherwise.
    (c) Effective description as a system in $Nd$ dimensions.
    The gray cube suggests the region where $\chi_{[0, 1]}\bm{(} \max_i \{ r_i(t) \} \bm{)} = 1$ in the ($Nd$)-dimensional space spanned by the orthogonal $N$ coordinates of the particles.
    Please refer to Fig.~\ref{fig:schematic_3d}(a) for comparison with the single-particle problem.}
    \label{fig:schematic_3n}
\end{figure}

\section{Temporal phase separation in multi-particle systems}
\label{sec:number}

As discussed in Sec.~\ref{sec:dimension}, a single Brownian particle system undergoes a DPT in five or higher dimensions.
This can be attributed to the general properties of eigenvalues of Schr\"{o}dinger-type equations~\cite{Klaus1980}.
For any $d \geq 2$, the $d$-dimensional Brownian motion can be considered as a combination of non-interacting one-dimensional Brownian motion.
Similarly, the eigenvalue problem for $N$ independent Brownian particles in $d$ dimensions is essentially equivalent to that for a single Brownian particle in $Nd$ dimensions.
According to the results in Sec.~\ref{sec:dimension}, the first-order DPT is expected to appear when $Nd \geq 5$.
We also expect that the situation for $d \to \infty$ in a single-particle setup (Sec.~\ref{sec_dim_asympt_law}) corresponds to the limit of $N \to \infty$ in the multi-particle setup, which can be considered as a macroscopic limit.

As an illustrative example, for $N$ non-interacting Brownian particles in $d$ dimensions, we consider the time fraction $\rho_T^{(N)}$ that \emph{all} the particles are in the $d$-dimensional unit ball centered at the origin, during the observation time $T$.
Writing the $i$th particle position as $\bm{x}_i(t)$ with $r_{i}(t) := ||\bm{x}_i(t)||$, we can express $\rho_T^{(N)}$ as
\begin{align}
    \rho_T^{(N)} &= \frac{1}{T} \int_0^T \left[ \prod_{i=1}^N \chi_{[0,1]}\bm{(}r_i(t)\bm{)} \right] dt
    \label{eq:obs_prod} \\
    &= \frac{1}{T} \int_0^T \chi_{[0,1]} \bm{(} \max_i \{ r_{i} (t) \} \bm{)} dt.
\end{align}
Note that $\rho_T^{(1)} = \rho_T$ [see Eq.~\eqref{eq:observable_rho}].
This expression suggests that $\rho_T^{(N)}$ is equivalent to the time fraction that a single particle in $N d$ dimensions is in a region near the origin, specified by the intersection of $N$ hypercylinders with $d$-dimensional spherical cross-sections [see Fig.~\ref{fig:schematic_3n}(c) for the case of $(N, d) = (3, 1)$].
Thus, the minimum number of particles $N_{\mathrm{min}}$ that satisfies the necessary condition for the first-order DPT (i.e., $Nd \geq 5$) depends on the spatial dimension: $N_{\mathrm{min}} = 5$ for $d = 1$, $N_{\mathrm{min}} = 3$ for $d = 2$, and $N_{\mathrm{min}} = 2$ for $d = 3$, for example.

\begin{figure}[t]
    \centering
    \includegraphics[scale=1]{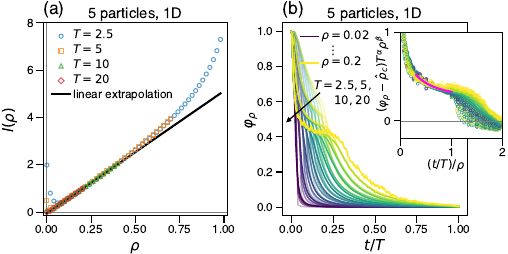}
    \caption{Phase separation in the multi-particle setup.
    (a) Rate function $I(\rho)$ for five particles in one dimension.
    Colored symbols are the data points obtained by simulations for several values of $T$.
    The black solid line is the linear extrapolation of two data points at $\rho = 0.1$ and $\rho = 0.2$ for $T = 20$, which suggests the existence of a linear region corresponding to phase separation.
    (b) Time dependence of the order parameter $\varphi_\rho$ for several values of $\rho$ and $T$.
    The inset shows the scaling plot with the parameter set $(\alpha, \beta, \hat{\rho}_c) \approx (0.38, 0.35, 0.21)$ estimated by fitting in the same way as used in Fig.~\ref{fig_dimensionality_ps}.}
    \label{fig_manybody_ps}
\end{figure}

To see the DPT in a multi-particle setup with the observable $\rho_T^{(N)}$, we conducted simulations of five non-interacting Brownian particles in one dimension [$(N, d) = (5, 1)$].
In Fig.~\ref{fig_manybody_ps}(a), we plot the rate function $I(\rho)$ for the obtained distribution $P(\rho_T^{(N)} = \rho)$ with colored symbols.
We also plot the linear line extrapolated from two data points at $\rho = 0.1$ and $\rho = 0.2$, which suggests a first-order DPT characterized by the existence of a linear region, $0 < \rho < {}^\exists \rho_c$.
As expected, this behavior of $I(\rho)$ is similar to that for a single particle in five dimensions [$(N, d) = (1, 5)$], which is plotted in Fig.~\ref{fig_dimensionality_ps}(i).

To numerically check the temporal phase separation associated with the first-order DPT, we introduce the order parameter similar to Eq.~\eqref{eq_order_param}, $\varphi_\rho (t) := \braket{\chi_{[0, 1]}\bm{(} \max_i \{ r_i(t) \} \bm{)}}$, which is shown in Fig.~\ref{fig_manybody_ps}(b).
We see the growth of a plateau-like region as $T$ increases, as observed for $(N, d) = (1, 5)$ [Fig.~\ref{fig_dimensionality_ps}(j)].
In the same way as used to obtain the asymptotic behavior when $T \to \infty$ for $(N, d) = (1, 5)$ [inset of Fig.~\ref{fig_dimensionality_ps}(j)], we tried fitting of $\varphi_\rho (t)$ for several values of $T$ and $\rho$.
We show the scaling plot with the optimal fitting parameters [$(\alpha, \beta, \hat{\rho}_c) \approx (0.38, 0.35, 0.21)$] in the inset of Fig.~\ref{fig_manybody_ps}(b).
Though the points look scattered compared to the case with $(N, d) = (1, 5)$ [inset of Fig.~\ref{fig_dimensionality_ps}(j)], this can be caused by finite-$T$ effects due to the anisotropic support of the integrand of $\rho_T^{(N)}$ [see Fig.~\ref{fig:schematic_3n}(b)].
We need simulations with larger $T$ to reduce the finite-$T$ effects and also to confirm if the obtained nominal critical point $\hat{\rho}_c$ ($\approx 0.21$) is close to the actual $\rho_c$.

We can extend the criterion for the first-order DPT to a combined system of multiple Brownian particles in different dimensions.
For example, we consider an $N$-particle system in $d$ dimensions and an $N'$-particle system in $d'$ dimensions.
After rescaling the length unit so that the diffusion constant is the same for both systems, the combined trajectory for the $(N + N')$ particles is equivalent to that for a single Brownian particle in $(Nd+N'd')$ dimensions.
Thus, for an observable that is defined similarly to Eq.~\eqref{eq:obs_prod}, the necessary condition for the first-order DPT is $Nd+N'd' \geq 5$.

The product form of $\rho_T^{(N)}$ [Eq.~\eqref{eq:obs_prod}] ensures that the integrand of the observable vanishes quickly as the distance from the origin increases in the equivalent single-particle problem in $Nd$ dimensions: $\prod_{i=1}^N \chi_{[0,1]} \bm{(} r_i \bm{)} = 0$ for $\sqrt{\sum_{i=1}^N {r_i}^2} > \sqrt{N}$.
This short-range property of the integrand seems essential to the effective increase in dimension from $d$ to $Nd$ in the large deviation of the observable.
For example, an observable in the form of summation~\cite{Mukherjee2023},
\begin{equation}
    \frac{1}{T} \int_0^T \left[ \sum_{i=1}^N \chi_{[0,1]}\bm{(}r_i(t)\bm{)} \right] dt,
\end{equation}
does not yield the effective high dimensionality; in this case, the integrand does not necessarily approach zero for $\sqrt{\sum_{i=1}^N {r_i}^2} \to \infty$.

\begin{table*}[t]
\centering
\begin{tabular}{c || c |  c |  c |  c |  c}
 & \multicolumn{2}{|c|}{Case (i): $\int_{-\infty}^{\infty} f(z) dz > 0$}  & Case (ii): $\int_{-\infty}^{\infty} f(z) dz = 0$ & \multicolumn{2}{|c}{Case (iii): $\int_{-\infty}^{\infty} f(z) dz < 0$} \\
\hline
SCGF $\lambda(k)$ for $\lambda(k) \approx 0$ & $ k^2$ & $(k-k_c)^2$ & $k^4$ & $(k-k_c)^2$ & $k^2$ \\
 & $( k > 0,\ k \approx 0)$ & $(k < k_c < 0,\ k \approx k_c)$ & $(k \approx 0)$ & $(k > k_c>0,\ k \approx k_c)$ & $(k < 0,\ k \approx 0)$ \\
\hline
Rate function $I(\sigma)$ for $\sigma \approx 0$ & $\sigma^2$ & $|\sigma|$ & $|\sigma|^{4/3}$ & $\sigma$ & $\sigma^2$  \\
 & $( \sigma > 0,\ \sigma \approx 0)$ & $(\sigma<0,\ \sigma \approx 0)$ & $(\sigma \approx 0)$ & $(\sigma>0,\ \sigma \approx 0)$ & $(\sigma < 0,\ \sigma \approx 0)$ \\
\end{tabular}
\caption{Power-law asymptotic forms of the SCGF $\lambda(k)$ and the rate function $I(\sigma)$ for three cases of the observable integrand $f(z)$ in one-dimensional Brownian motion.
We omit the constants of proportionality.}
\label{tab_expanded_scgf_1d}
\end{table*}

\begin{figure}[t]
    \centering
    \includegraphics[scale=1]{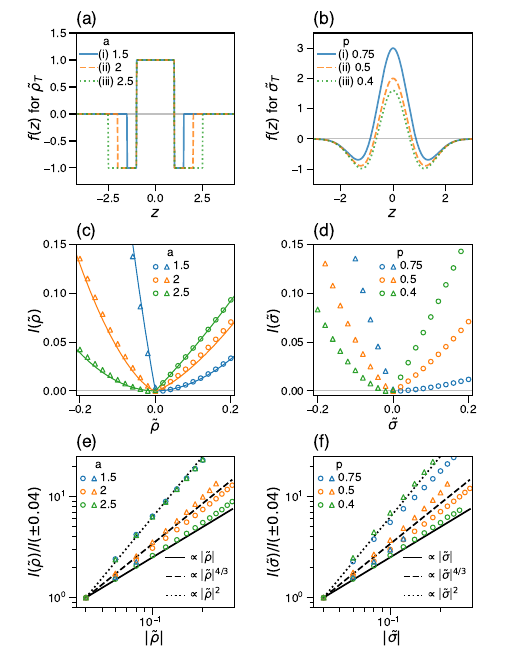}
    \caption{Universal exponents of the rate functions in one-dimensional Brownian motion.
    (a, b) Integrands $f(z)$ of the observables $\tilde{\rho}_T$ [Eq.~\eqref{eq:tilde_rho}] and $\tilde{\sigma}_T$ [Eq.~\eqref{eq:tilde_sigma}].
    In (a), we plot $f(z)$ for (i) $a = 1.5$ (blue solid line), (ii) $a = 2$ (orange dashed line), and (iii) $a = 2.5$ (green dotted line).
    In (b), we plot $f(z)$ for (i) $p = 0.75$ (blue solid line), (ii) $p = 0.5$ (orange dashed line), and (iii) $p = 0.4$ (green dotted line).
    These cases (i)-(iii) correspond to $\int_{-\infty}^{\infty} f(z)dz > 0$, $\int_{-\infty}^{\infty} f(z)dz = 0$, and $\int_{-\infty}^{\infty} f(z)dz < 0$, respectively.
    (c, d) Rate functions of $\tilde{\rho}$ and $\tilde{\sigma}$, whose integrands are shown in (a) and (b), respectively.
    The colored symbols are the simulation results, and the solid lines in (c) are the theoretical predictions.
    The data suggest non-differentiability at (c) $\tilde{\rho} = 0$ or (d) $\tilde{\sigma} = 0$ for cases (i) and (iii).
    Parameters are set to $dt = 0.05$, $T = 80$ for simulations and we take $10^8$ samples to calculate the rate functions.
    The number of trajectories are counted with bin size of $\tilde{\rho}$ and $\tilde{\sigma}$ being $0.02$ and are normalized so that $I(\tilde{\rho})$ and $I(\tilde{\sigma})$ are $0$ at their minima.
    (e, f) Log-log plot of (c) and (d) with the horizontal axis taken as $|\tilde{\rho}|$ and $|\tilde{\sigma}|$, respectively.
    The black lines are power functions with the theoretically predicted exponents, which agree well with the observations.}
    \label{fig3_23}
\end{figure}

\section{Universal localization transition in one dimension}
\label{sec:universality}

The results in Sec.~\ref{sec:dimension} suggest that no DPT appears in the dynamical path in one-dimensional Brownian motion as long as we take the simple time-averaged observable defined by Eq.~\eqref{eq:observable_rho}.
In this section, we search for possible DPTs and their universal properties in one dimension ($d = 1$) by generalizing the observable [Eq.~\eqref{eq_observable}]:
\begin{equation}
    \sigma_T = \frac{1}{T} \int^T_0 f\bm{(}x(t)\bm{)} dt.
    \label{eq:obs_1d}
\end{equation}
Here, we assume that $f(z)$ is a real scalar function with $\int^{\infty}_{-\infty} (1+|z|)|f(z)| dz < \infty$~\cite{Klaus1977} and has a sign change [i.e., $\max_z{\{ f(z) \}} \min_z{\{ f(z) \}} < 0$]
~\footnote{If $f(z)$ is a non-negative function, $\lambda (k)$ is zero for $k < 0$ and quadratic for $k > 0$, and thus $\sigma$ only takes non-negative values, as exemplified in Sec.~\ref{sec:derivation_1d}. If $f(z)$ is a non-positive function, $\lambda (k)$ is zero for $k > 0$ and quadratic for $k < 0$, and thus $\sigma$ only takes non-positive values. In both cases, $I(\sigma) \propto \sigma^2$ and differentiable everywhere.}.
Note that $\sigma_T$ can be negative depending on the path $x(t)$.

\subsection{Second-order DPTs in SCGF}

To obtain the SCGF $\lambda (k)$, we consider the eigenvalue problem [Eq.~\eqref{eq:spectrum}]:
\begin{equation}
    \mathcal{L}_k \phi_k(x) = \lambda(k) \phi_k(x),
    \label{eq_eigen_1d}
\end{equation}
where $\mathcal{L}_k = d^2 / dx^2 + kf(x)$.
In the same way as explained in Sec.~\ref{sec:dimension}, if $\lambda (k) > 0$, the wavefunction $\phi_k (x)$ is a localized (i.e., bound) state around a certain region.

According to Refs.~\cite{Klaus1980,Simon1976,Klaus1977,Bronzan1987}, we obtain the following asymptotic behaviors of $\lambda (k)$, depending on the functional form of $f(z)$.
We obtain the following three distinct cases:

\textbf{Case (i): $\int_{-\infty}^{\infty} f(z)dz > 0$}
\begin{equation}
    \lambda(k) 
    \left\{ \,
    \begin{aligned}
    & \approx ck^2 & (k > 0,\ k \approx 0) \\
    & = 0 & (k_c < k < 0) \\
    & \approx c'(k-k_c)^2 & (k < k_c,\ k \approx k_c)
    \end{aligned}
    \right. .
    \label{eq_case_i_lam}
\end{equation}

\textbf{Case (ii): $\int_{-\infty}^{\infty} f(z)dz = 0$}
\begin{equation}
    \lambda(k) \approx c k^4 \ \ \ (k \approx 0) .
    \label{eq_case_ii_lam}
\end{equation}

\textbf{Case (iii): $\int_{-\infty}^{\infty} f(z)dz < 0$}
\begin{equation}
    \lambda(k) 
    \left\{ \,
    \begin{aligned}
    & \approx c(k-k_c)^2 & (k > k_c,\ k \approx k_c) \\
    & = 0 & (0 < k < k_c) \\
    & \approx c'k^2 & (k < 0,\ k \approx 0)
    \end{aligned}
    \right. .
    \label{eq_case_iii_lam}
\end{equation}
Here, $c$ and $c'$ ($ > 0$) are determined by the specific form of $f(z)$, and $k_c$ is a negative (positive) threshold for the localized state formation when $\int_{-\infty}^{\infty} f(z)dz > 0$ ($< 0$).
The asymptotic behavior of $\lambda(k)$ is summarized in Table~\ref{tab_expanded_scgf_1d}.

For cases (i) [Eq.~\eqref{eq_case_i_lam}] and (iii) [Eq.~\eqref{eq_case_iii_lam}], we find that the second derivative of $\lambda (k)$ is discontinuous at $k = 0$ and $k = k_c$, indicating two sequential second-order DPTs in the SCGF.
Through the DPTs, two kinds of localized states with $\lambda (k) > 0$ and a non-localized state with $\lambda (k) = 0$ appear.
Note that similar sequential DPTs in the SCGF have been found in the kinetic Ising model~\cite{Jack2010}, where energy is taken as the observable, and three magnetically distinct phases appear through DPTs.

\subsection{Power-law rate function and localization transition}
\label{subsec_univ_pow_rate_func}

By the Legendre-Fenchel transformation [Eq.~\eqref{eq:L-F}], we obtain the rate function $I(\sigma)$ for cases (i)-(iii) above (see Appendix~\ref{sec_LF} for the derivations).
We first focus on case (i), where $\int_{-\infty}^{\infty} f(z)dz > 0$.
From the asymptotic forms of $\lambda(k)$ in Eq.~\eqref{eq_case_i_lam}, we obtain the asymptotic forms of $I(\sigma)$ for $\sigma \approx 0$.

\textbf{Case (i): $\int_{-\infty}^{\infty} f(z)dz > 0$}
\begin{equation}
    I(\sigma) \approx
    \left\{ \,
    \begin{aligned}
    & \frac{\sigma^2}{4c} & (\sigma > 0¥,\ \sigma \approx 0) \\
    & |k_c \sigma| & (\sigma < 0,\ \sigma \approx 0) 
    \end{aligned}
    \right. .
    \label{eq_case_i_I}
\end{equation}
Thus, the exponents $\alpha_{\pm}$ defined from the power law $I(\sigma) \propto |\sigma|^{\alpha_{\pm}}$ for $\sigma \gtrless 0$ are universal ($\alpha_+ = 2$ and $\alpha_- = 1$), i.e., independent of the specific form of $f(z)$.

\textbf{Case (ii): $\int_{-\infty}^{\infty} f(z)dz = 0$}
\begin{equation}
    I(\sigma) \approx \frac{3c |\sigma|^{4/3}}{(4c)^{4/3}} \ \ \ (\sigma \approx 0) .
    \label{eq_case_ii_I}
\end{equation}
The universal exponents are $\alpha_+ = \alpha_- = 4/3$.

\textbf{Case (iii): $\int_{-\infty}^{\infty} f(z)dz < 0$}
\begin{equation}
    I(\sigma) \approx
    \left\{ \,
    \begin{aligned}
    & k_c \sigma & (\sigma > 0,\ \sigma \approx 0) \\
    & \frac{\sigma^2}{4c'} & (\sigma < 0,\ \sigma \approx 0)
    \end{aligned}
    \right. .
    \label{eq_case_iii_I}
\end{equation}
The two exponents $\alpha_{\pm}$ are opposite to those for case (i): $\alpha_+ = 1$ and $\alpha_- = 2$.
The power-law behavior of $I(\sigma)$ is summarized in Table~\ref{tab_expanded_scgf_1d}.

For cases (i) and (iii), $I(\sigma)$ is not differentiable at $\sigma = 0$, which suggests that dynamical paths conditioned by $\sigma_T = \sigma$ qualitatively change when crossing $\sigma = 0$.
This is regarded as a localization transition since $\sigma > 0$ or $\sigma < 0$ represents atypical paths $x(t)$ that localize around the region where $f \bm{(} x(t) \bm{)} > 0$ or $f \bm{(} x(t) \bm{)} < 0$, respectively.
Consistently, through the Legendre-Fenchel transformation, $\sigma \neq 0$ leads to $\lambda (k) > 0$, whose eigenfunction $\phi_k(x)$ is localized as well as the conditional distribution $P(x | \sigma_T = \sigma) \propto \phi_k (x)^2$ (see Sec.~\ref{sec:formulation}).
Note that $\sigma = 0$ represents typical paths that show non-localized normal diffusion, as also suggested by $\lambda (k) = 0$.

\subsection{Examples of localization transition}

We demonstrate that two different observables show the localization transition with the power-law rate function as predicted in Sec.~\ref{subsec_univ_pow_rate_func}.

\subsubsection{Example 1: difference between two occupations}

As the first example, we consider the difference between the fraction of time satisfying $|x(t)| \in [0, 1]$ and the fraction of time satisfying $|x(t)| \in [1, a]$:
\begin{equation}
    \tilde{\rho}_T := \frac{1}{T} \int^T_0 \chi_{[0,1]}\bm{(}|x(t)|\bm{)} dt - \frac{1}{T} \int^T_0 \chi_{[1,a]}\bm{(}|x(t)|\bm{)} dt.
    \label{eq:observable_diff}
\end{equation}
We set $x(0) = 0$ as the initial condition.
In Fig.~\ref{fig3_23}(a), we plot the integrand of $\tilde{\rho}_T$:
\begin{equation}
    f(z) = \chi_{[0,1]}(|z|) - \chi_{[1,a]}(|z|).
    \label{eq:tilde_rho}
\end{equation}
From $\int_{-\infty}^{\infty} f(z)dz = 2(2-a)$, we chose (i) $a=1.5$, (ii) $a = 2$, and (iii) $a = 2.5$ as representative parameters for $\int_{-\infty}^{\infty} f(z)dz > 0$, $\int_{-\infty}^{\infty} f(z)dz = 0$, and $\int_{-\infty}^{\infty} f(z)dz < 0$, respectively.

In Fig.~\ref{fig3_23}(c), we show the rate function $I(\tilde{\rho})$ obtained from simulations (colored symbols).
For (i) $a=1.5$ and (iii) $a = 2.5$, $I(\tilde{\rho})$ is non-differentiable at $\tilde{\rho} = 0$, which is more clearly seen in the log-log plot of $I(\tilde{\rho})$ vs $|\tilde{\rho}|$ [Fig.~\ref{fig3_23}(e)].
As predicted from the general theory for cases (i)-(iii) [Eqs.~\eqref{eq_case_i_I}-\eqref{eq_case_iii_I}], we asymptotically obtain the power law $I(\tilde{\rho}) \sim |\tilde{\rho}|^{\alpha_{\pm}}$ for $\tilde{\rho} \gtrless 0$ with (i) $\alpha_+ = 2$ (blue circle) and $\alpha_- = 1$ (blue triangle), (ii) $\alpha_+ = 4/3$ (orange circle) and $\alpha_- = 4/3$ (orange triangle), and (iii) $\alpha_+ = 1$ (green circle) and $\alpha_- = 2$ (green triangle).

Solving the eigenvalue problem for the dominant eigenfunction $\phi_k (x)$ [Eq.~\eqref{eq_eigen_1d}] with $\mathcal{L}_k = d^2/dx^2 + k[\chi_{[0,1]}(|x|)-\chi_{[1,a]}(|x|)]$, we can analytically obtain the SCGF $\lambda(k)$ and the rate function $I(\tilde{\rho})$ for $T \to \infty$, in a similar way as used in Sec.~\ref{sec:dimension}.
Imposing the convergence of $\phi_k (x)$ for $x \to \pm \infty$ and using $\phi_k(-x) = \phi_k(x)$ since $f(z)$ is an even function, we obtain
\begin{equation}
    \phi_k(x) = 
    \left\{ \,
    \begin{aligned}
    & C_k\cos{(\zeta_k |x|)} & (0<|x|<1) \\
    & D_k\cosh{(\eta_k |x|)}+E_k\sinh{(\eta_k |x|)} & (1<|x|<a) \\
    & F_k\exp{(-\gamma_k |x|)} & (a<|x|)
    \end{aligned}
    \right. ,
    \label{eq_ex1_phik_pos}
\end{equation}
for $k>0$, and
\begin{equation}
    \phi_k(x)=
    \left\{ \,
    \begin{aligned}
    & C'_k\cosh{(\eta_k |x|)} & (0<|x|<1) \\
    & D'_k\cos{(\zeta_k |x|)}+E'_k\sin{(\zeta_k |x|)} & (1<|x|<a) \\
    & F'_k\exp{(-\gamma_k |x|)} & (a<|x|)
    \end{aligned}
    \right. ,
    \label{eq_ex1_phik_neg}
\end{equation}
for $k<0$.
Here, the coefficients $\{ C_k^{(\prime)}, D_k^{(\prime)}, E_k^{(\prime)}, F_k^{(\prime)} \}$ and the eigenvalue $\lambda (k)$ are determined by the boundary conditions [Eqs.~\eqref{eq:SCGF-p} and \eqref{eq:SCGF-m}] and the normalization condition.
Also, $\gamma_k := \sqrt{\lambda(k)}$, $\zeta_k := \sqrt{|k|-\lambda(k)}$, and $\eta_k := \sqrt{|k|+\lambda(k)}$.

$\lambda(k)$ is given as the solution of the following equations obtained from the boundary conditions at $|x|=1$ and $|x| = a$:
\begin{equation}
    \zeta_k \tan{(\zeta_k)} = \eta_k \frac{\gamma_k + \eta_k \tanh{\bm{(}\eta_k(a-1)\bm{)}}}{\eta_k + \gamma_k \tanh{\bm{(}\eta_k(a-1)\bm{)}}} \ \ \ \text{for} \ k>0, 
    \label{eq:SCGF-p}
\end{equation}
and
\begin{equation}
    - \eta_k \tanh{(\eta_k)} = \zeta_k \frac{\gamma_k - \zeta_k \tan{\bm{(}\zeta_k (a-1)\bm{)}}}{\zeta_k + \gamma_k \tan{\bm{(}\zeta_k (a-1)\bm{)}}} \ \ \ \text{for} \ k<0.
    \label{eq:SCGF-m}
\end{equation}
For small $|k|$, one can show that depending on $a$, SCGF $\lambda(k)$ behaves as:
\begin{equation}
    \lambda(k)
    \left\{\,
    \begin{aligned}
    &\approx (2-a)^2 k^2 & (k > 0 ,\ k \approx 0) \\
    &= 0 & (k_c < k < 0) \\
    &\approx c_a (k-k_c)^2 &(k < 0,\ k \approx k_c) \\
    \end{aligned}
    \right.
    \ \ \ \text{for} \ a<2,
\end{equation}
\begin{equation}
    \lambda(k) \approx \frac{a^2}{9}|k|^4 = \frac{4}{9}|k|^4 \ \ \ (k \approx 0) \ \ \ \text{for}  \ a=2,
\end{equation}
and
\begin{equation}
    \lambda(k)
    \left\{\,
    \begin{aligned}
    &\approx c_a (k-k_c)^2 & (k > k_c,\ k \approx k_c) \\
    &= 0 & (0 < k < k_c) \\
    &\approx (a-2)^2 k^2 & (k < 0,\ k \approx 0) \\
    \end{aligned}
    \right.
    \ \ \ \text{for}  \ a>2,
    \label{eq_ex1_case_iii}
\end{equation}
with constant $c_a$ and nonzero $k_c$.
Equations to determine $k_c$ are
\begin{equation}
    \tan{\left( \sqrt{k_c} \right)} = \tanh{\left( \sqrt{k_c}(a-1) \right)} \ \ \ \text{for} \ k>0
\end{equation}
and
\begin{equation}
    \tanh{\left( \sqrt{-k_c} \right)} = \tan{\left( \sqrt{-k_c}(a-1) \right)} \ \ \ \text{for} \ k<0.
\end{equation}
Thus, for small $|\tilde{\rho}|$,
\begin{equation}
    I(\tilde{\rho}) \approx
    \left\{\,
    \begin{aligned}
    &\frac{1}{4(2-a)^2} \tilde{\rho}^2 & (\tilde{\rho} > 0,\ \tilde{\rho} \approx 0) \\
    &|k_c \tilde{\rho}| &(\tilde{\rho} < 0,\ \tilde{\rho} \approx 0) \\
    \end{aligned}
    \right.
    \ \ \ \text{for} \ a<2,
\end{equation}
\begin{equation}
    I(\tilde{\rho}) \approx \left( \frac{3^5}{2^8a^2} \right)^{1/3} |\tilde{\rho}|^{4/3} = \left( \frac{3}{4} \right)^{5/3} |\tilde{\rho}|^{4/3} \ \ \ (\tilde{\rho} \approx 0) \ \ \ \text{for}  \ a=2,
\end{equation}
and
\begin{equation}
    I(\tilde{\rho}) \approx
    \left\{\,
    \begin{aligned}
    &k_c \tilde{\rho} & (\tilde{\rho} > 0,\ \tilde{\rho} \approx 0) \\
    &\frac{1}{4(a-2)^2} \tilde{\rho}^2 & (\tilde{\rho} < 0,\ \tilde{\rho} \approx 0) \\
    \end{aligned}
    \right.
    \ \ \ \text{for}  \ a>2.
\end{equation}
The results are consistent with the general behavior shown in Table~\ref{tab_expanded_scgf_1d}.
The obtained $I(\tilde{\rho})$ for $T \to \infty$ is plotted as colored solid lines in Fig.~\ref{fig3_23}(c), showing quite good agreement with the data from simulations for all the cases (i)-(iii) (colored symbols).

\begin{figure}[t]
    \centering
    \includegraphics[scale=1]{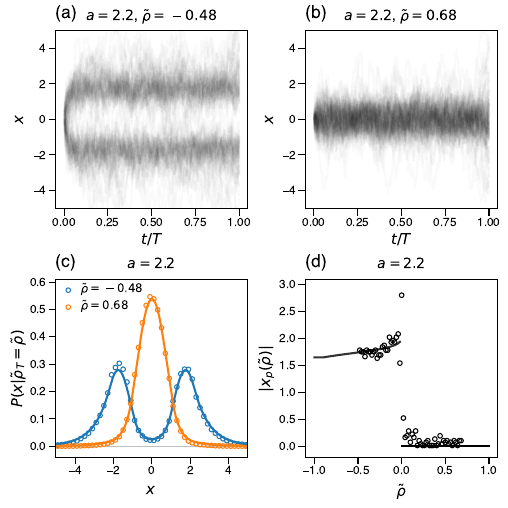}
    \caption{Localization transition as DPTs in one-dimensional Brownian motion.
    (a, b) 100 typical trajectories of the particle conditioned with (a) $\tilde{\rho} = -0.48$ and (b) $\tilde{\rho} = 0.68$.
    (c) Conditional probability density $P(x|\tilde{\rho}_T = \tilde{\rho})$ for $\tilde{\rho} = -0.48$ and $\tilde{\rho} = 0.68$, corresponding to (a) and (b), respectively.
    The colored circles and solid lines are obtained from simulations and the analytical expression [i.e., $\phi_{k^*(\tilde{\rho})}(x)^2$ with Eqs.~\eqref{eq_ex1_phik_pos} and \eqref{eq_ex1_phik_neg}], respectively, showing good agreement.
    The numerical $P(x|\tilde{\rho}_T=\tilde{\rho})$ is calculated from $10^3$ independent trajectories in the time domain of $1/4 < t/T < 3/4$ with the bin size of $\tilde{\rho}$ being $0.02$.
    (d) $\tilde{\rho}$ dependence of the peak position of $P(x|\tilde{\rho}_T = \tilde{\rho})$, $|x_p(\tilde{\rho})|$, which shows a jump at $\tilde{\rho} = 0$.
    The circles and solid lines are obtained from simulations and the analytical expression [Eq.~\eqref{eq:peak}], respectively.
    The numerical $|x_p(\tilde{\rho})|$ is calculated from $10^3$ independent trajectories in the time domain of $1/4 < t/T < 3/4$ for each $\tilde{\rho}$ with the bin sizes of $\tilde{\rho}$ and $|x|$ being $0.02$ and $0.03$, respectively.
    Parameters for (a-d): $a = 2.2$, 
    $dt = 0.05$, and $T = 20$.}
    \label{fig2_22}
\end{figure}

To illustrate the difference in localization between $\tilde{\rho} < 0$ and $\tilde{\rho} > 0$, we focus on case $a > 2$ (specifically, $a = 2.2$).
In Figs.~\ref{fig2_22}(a) and (b), we plot the dynamical path $x(t)$ conditioned with $\tilde{\rho} = -0.48$ ($< 0$) and $\tilde{\rho} = 0.68$ ($> 0$), respectively.
As predicted in Sec.~\ref{subsec_univ_pow_rate_func}, $x(t)$ is typically localized around the region with $f\bm{(} x(t) \bm{)} < 0$ [i.e., $1 < |x(t)| < 2.2$] for $\tilde{\rho} < 0$; in contrast, $x(t)$ is localized around the region with $f\bm{(} x(t) \bm{)} > 0$ [i.e., $0 < |x(t)| < 1$] for $\tilde{\rho} > 0$.
We plot the conditional distribution $P(x | \tilde{\rho}_T = \tilde{\rho})$ for each $\tilde{\rho}$ in Fig.~\ref{fig2_22}(c) (colored circles), which also visualizes the difference in localization and is consistent with the theoretical prediction for $T \to \infty$ (colored lines): $P(x | \tilde{\rho}_T = \tilde{\rho}) \propto \phi_k (x)^2$ (see Sec.~\ref{sec:formulation}) with $\phi_k (x)$ calculated from Eqs.~\eqref{eq_ex1_phik_pos} and \eqref{eq_ex1_phik_neg}.

The localization transition at 
$\tilde{\rho} = 0$ should be characterized by the change in the localization center of the dynamical path $x(t)$.
To examine this, we consider the $\tilde{\rho}$ dependence of $x_p (\tilde{\rho})$, the peak position(s) of $P(x | \tilde{\rho}_T = \tilde{\rho})$. 
As shown with black circles in Fig.~\ref{fig2_22}(d), we find that $|x_p (\tilde{\rho})|$ obtained by simulations shows a jump at the transition point ($\tilde{\rho} = 0$).
Based on the peak position(s) of $P(x | \tilde{\rho}_T = \tilde{\rho}) \propto \phi_k (x)^2$ with $\phi_k (x)$ calculated from Eqs.~\eqref{eq_ex1_phik_pos} and \eqref{eq_ex1_phik_neg}, we also obtain the asymptotic form of $x_p (\tilde{\rho})$ for $T \to \infty$:
\begin{equation}
    x_p (\tilde{\rho}) =
    \left\{ \,
    \begin{aligned}
    &0 & (k>0) \\
    &\pm \frac{1}{\zeta_{k^*(\tilde{\rho})}}\arctan{\left( \frac{E_{k^*(\tilde{\rho})}'}{D_{k^*(\tilde{\rho})}'} \right)} \ (\neq 0) & (k<0)
    \end{aligned}
    \right. ,
    \label{eq:peak}
\end{equation}
where $k^*(\tilde{\rho})$ is determined by $k^*(\tilde{\rho}) = \mathrm{argmax}_k \{k \tilde{\rho} - \lambda (k) \}$ (see Sec.~\ref{sec:formulation}).
In Fig.~\ref{fig2_22}(d), we plot this result with black lines, which shows a good agreement with the data.
Thus, the localization changes abruptly as $|x_p (\tilde{\rho})| > 0$ for $\tilde{\rho} < 0$ and $x_p (\tilde{\rho}) = 0$ for $\tilde{\rho} > 0$.

\subsubsection{Example 2: integral of a smooth function with sign change}

As the second example to support the predicted universality, we consider an observable $\tilde{\sigma}_T = T^{-1} \int_0^T f\bm{(}x(t)\bm{)} dt$ with
\begin{equation}
    f(z)=-4e^{-z^2}(z^2-p).
    \label{eq:tilde_sigma}
\end{equation}
Here, the parameter $p$ controls the balance between the positive and negative regions of $f(z)$, as suggested by
\begin{equation}
    \int_{-\infty}^{\infty} f(z) dz = -4 \sqrt{\pi} \left( \frac{1}{2} - p \right).
\end{equation}
In Fig.~\ref{fig3_23}(b), we plot the functional form of $f(z)$ for (i) $p = 0.75$ ($\int_{-\infty}^{\infty} f(z)dz > 0$), (ii) $p = 0.5$ ($\int_{-\infty}^{\infty} f(z)dz = 0$), and (iii) $p = 0.4$ ($\int_{-\infty}^{\infty} f(z)dz < 0$).

In Fig.~\ref{fig3_23}(d), we plot the rate function $I(\tilde{\sigma})$ obtained from simulations, which looks non-differentiable at $\tilde{\sigma} = 0$ for (i) $p = 0.75$ and (iii) $p = 0.4$.
Figure~\ref{fig3_23}(f) is the corresponding log-log plot, where we also plot power functions with exponents suggested in the legend.
For cases (i)-(iii), we find the asymptotic power law $I(\tilde{\sigma}) \sim |\tilde{\sigma}|^{\alpha_{\pm}}$ for $\tilde{\sigma} \gtrless 0$ with (i) $\alpha_+ = 2$ (blue circle) and $\alpha_- = 1$ (blue triangle), (ii) $\alpha_+ = 4/3$ (orange circle) and $\alpha_- = 4/3$ (orange triangle), and (iii) $\alpha_+ = 1$ (green circle) and $\alpha_- = 2$ (green triangle).
These behaviors are consistent with the general prediction [Eqs.~\eqref{eq_case_i_I}-\eqref{eq_case_iii_I}] and share the universal exponents with the previous example [see Fig.~\ref{fig3_23}(e)].

\section{Discussion and outlook}
\label{sec:conclusion}

We conclude by highlighting open questions and future research directions stemming from our findings.

While our observation of temporal phase separation associated with the first-order DPT echoes phenomena reported in Refs.~\cite{Nyawo2017,Nyawo2018}, we currently lack a theoretical understanding of why this DPT leads to temporal phase separation.
We establish the empirical asymptotics of the order parameter as a function of total time ($T$) and occupation ($\rho$), characterized by exponents $\alpha$ and $\beta$. We anticipate that these findings will provide clues for elucidating the mechanism underlying the observed temporal phase separation.

A key open question is the unexpected critical dimension of $d = 4$, in contrast to the typical $d = 2$ observed in diffusion-related problems due to the recurrent nature of Brownian motion. While the critical dimension $d = 4$ can be mathematically attributed to the change in Bessel function properties at $\nu = (d-2)/2 = 1$, a more intuitive understanding remains to be developed.
Based on the general arguments in Ref.~\cite{Klaus1980}, we anticipate that the first-order DPT will persist universally in dimensions higher than four regardless of the specific observable $f(\bm{x})$, provided it decays sufficiently rapidly as $||\bm{x}|| \to \infty$.
Interestingly, it is known that there is a transition in the statistics of random walk intersections at $d = 4$ ~\cite{Chen2010,Lawler2013}; the number of self-intersections grows to infinity in the limit of a long trajectory only when $d\leq 4$, owing to the Galiardo-Nirenberg-Sobolev inequality.
While a direct connection to our findings remains unclear, this intriguing coincidence suggests a potential avenue for investigation.

Our results indicate that the phase transition and its associated exponents exhibit universality for observables $f(\bm{x})$ that decay rapidly as $||\bm{x}|| \to \infty$, corresponding to short-range potentials in analogous quantum systems. 
Previous studies~\cite{Klaus1980, Bronzan1987} suggest that the coupling constant threshold can be changed when we consider long-range potentials [i.e., $f(\bm{x})$ that converges to zero slowly as $||\bm{x}|| \to \infty$].
This implies that other universality classes for the large deviation of the observables may appear for more slowly decaying $f(\bm{x})$.

\begin{acknowledgments}
We thank Takahiro Nemoto and Akira Shimizu for fruitful discussions.
This work was supported by JSPS KAKENHI Grant Numbers JP20K14435 (to K.A.), JP19H05795, JP19H05275, JP21H01007, and JP23H00095 (to K.K.).
\end{acknowledgments}

\appendix

\section{Legendre-Fenchel transformation of power functions}
\label{sec_LF}

We consider the following convex power functions with $a>0$ and $\beta \geq 1$:
\begin{equation}
    \textbf{Case (I)}\text{:} \ \ \ f(x) = a |x|^\beta,
    \label{eq:c0}
\end{equation}
\begin{align}
    \textbf{Case (II)}\text{:} \ \ \ 
    f(x) &=
    \left\{ \,
    \begin{aligned}
    &a (x-c)^\beta & (x>c) \\
    &0  & (x \leq c)
    \end{aligned}
    \right.
    \ \text{with} \ c > 0,
    \label{eq_app_case_II}
\end{align}
and
\begin{align}
    \textbf{Case (III)}\text{:} \ \ \ 
    f(x) &=
    \left\{ \,
    \begin{aligned}
    &0 & (x>c) \\
    &a |x-c|^\beta & (x \leq c)
    \end{aligned}
    \right.
    \ \text{with} \ c < 0.
    \label{eq_app_case_III}
\end{align}

The Legendre-Fenchel transformation of function $f(x)$, $g(X)$, is given by
\begin{equation}
    g(X)=\sup_x \{xX-f(x)\},
    \label{eq:LF}
\end{equation}
where $X$ is conjugated to $x$.
If $f(x)$ is convex, Eq.~\eqref{eq:LF} can be written as
\begin{equation}
    g(X) = x^* X - f(x^*),
\end{equation}
where $x^*$ satisfies
\begin{equation}
    f'(x^*-0) \leq X \leq f'(x^*+0)
    \label{eq:x_star_app}
\end{equation}
with $f'(x) := d f (x) / d x$.
When $f(x)$ is differentiable, Eq.~\eqref{eq:x_star_app} can also be written as
\begin{equation}
    X = f'(x^*).
    \label{eq:x_star_differentiable}
\end{equation}

First, we consider $g(X)$ for case (I).
When $\beta=1$, noticing that $f(x)$ ($= a |x|$) is undifferentiable at $x = 0$, we find that $g(X)$ is a constant function: $g(X)=0$ for $-a < X < a$.
When $\beta > 1$, we first consider the case with $x^* > 0$, where $X > 0$ follows from $f(x) = a x^\beta$ and Eq.~\eqref{eq:x_star_app}.
Since $X = f'(x^*) = a \beta {x^*}^{\beta - 1}$, we obtain $x^* = (X / a\beta)^{1 / (\beta - 1)}$, leading to
\begin{equation}
    g(X) = a(\beta-1) \left(\frac{1}{a \beta} \right)^{\frac{\beta}{\beta - 1}} X^{\frac{\beta}{\beta - 1}} \ \ \ (X>0).
\end{equation}
In a similar way, since $X = -a\beta(-x^*)^{\beta-1}$ for $x^*<0$, we obtain
\begin{equation}
    g(X) = a(\beta-1) \left(\frac{1}{a \beta} \right)^{\frac{\beta}{\beta - 1}} (-X)^{\frac{\beta}{\beta - 1}} \ \ \ (X<0).
\end{equation}

Next, we consider case (II).
When $\beta = 1$, noticing that $f(x)$ is undifferentiable at $x = c$ ($> 0$) [see Eq.~\eqref{eq_app_case_II}], we find that $g(X)$ is a linear function: $g(X) = cX$ for $0 < X < a$.
When $\beta > 1$, since $f'(x)=a \beta (x-c)^{\beta - 1}$ for $x > c$ and $f'(x) = 0$ for $x \leq c$, we obtain $x^* = (X / a \beta)^{1 / (\beta - 1)} + c$ and thus,
\begin{equation}
    g(X) = cX + a(\beta - 1) \left(\frac{1}{a \beta}\right)^{\frac{\beta}{\beta-1}} X^{\frac{\beta}{\beta-1}} \ \ \ (X > 0),
\end{equation}
which can be approximated as $g(X) \approx c X$ for small $X$.
We can apply the obtained results to case (III) by replacing $c$ and $X$ with $-c$ and $-X$, respectively, according to the functional form of $f(x)$ [see Eqs.~\eqref{eq_app_case_II} and \eqref{eq_app_case_III}].

In summary, we obtain the Legendre-Fenchel transformation of $f(x)$ for cases (I)-(III) [Eqs.~\eqref{eq:c0}-\eqref{eq_app_case_III}]:

\textbf{Case (I)}
\begin{equation}
    g(X) = 
    \left\{ \,
    \begin{aligned}
    &0 & (-a < X < a, \ \beta = 1) \\
    &a(\beta-1) \left(\frac{1}{a \beta} \right)^{\frac{\beta}{\beta - 1}} |X|^{\frac{\beta}{\beta - 1}} & (\beta > 1)
    \end{aligned}
    \right. .
\end{equation}

\textbf{Case (II)}
\begin{equation}
    g(X) =
    \left\{ \,
    \begin{aligned}
    &cX & (0 < X < a, \ \beta = 1) \\
    &cX + a(\beta - 1) \left(\frac{1}{a \beta}\right)^{\frac{\beta}{\beta-1}} X^{\frac{\beta}{\beta-1}} & (X > 0, \ \beta>1)
    \end{aligned}
    \right. .
\end{equation}

\textbf{Case (III)}
\begin{equation}
    g(X) =
    \left\{ \,
    \begin{aligned}
    &cX & (-a < X < 0, \ \beta = 1) \\
    &cX + a(\beta - 1) \left(\frac{1}{a \beta}\right)^{\frac{\beta}{\beta-1}} (-X)^{\frac{\beta}{\beta-1}} & (X < 0, \ \beta>1)
    \end{aligned}
    \right. .
\end{equation}

\end{document}